\documentclass[preprint,12pt,3p]{elsarticle}

\usepackage{graphicx}
\usepackage{caption}
\usepackage{subcaption}
\usepackage[unicode=true,
 bookmarks=true,bookmarksnumbered=false,bookmarksopen=false,
 breaklinks=false,pdfborder={0 0 1},colorlinks=true]{hyperref}
\usepackage{amssymb,amsmath}
\usepackage{wasysym}
\usepackage{float}
\usepackage{amsthm}
\usepackage{dsfont}
\usepackage{tikz}
\usepackage{tkz-euclide}
\usepackage{adjustbox}
\usepackage{bbold}
\usepackage{braket}
\usepackage{natbib}
\usepackage{lineno}

\usepackage{tikz}

\journal{Annals of Physics}

\begin{document}

	\begin{frontmatter}
		\title{Non-Abelian fusion rules from Abelian systems\\ with SPT phases and graph topological order}
	
		\author[usp]{M. F. Araujo de Resende\corref{cor1}}
		\ead{resende@if.usp.br}
		\cortext[cor1]{Corresponding author}
		\author[latam]{J. P. Ibieta Jimenez}
		\ead{ibieta.pablo@gmail.com}
		\author[temuco]{J. Lorca Espiro}
		\ead{javier.lorca@ufrontera.cl}
	
		\address[usp]{Instituto de F\'{\i}sica, Universidade de S\~{a}o Paulo, 05508-090 S\~{a}o Paulo SP, Brasil}
		
		\address[latam]{Experian DataLab LatAm, 04794-000 São Paulo SP, Brazil}
	
		\address[temuco]{Departamento de Ciencias F\'{\i}sicas, Facultad de Ingenier\'{\i}a,
Ciencias y Administraci\'{o}n, Universidad de La Frontera, Avda. Francisco Salazar 01145, Casilla 54-D Temuco, Chile}
	
		\begin{abstract}
			Since Ref. \cite{pramod} shows the emergence of non-Abelian fusion rules in some examples of a class of Abelian models, but does not prove whether these rules also exist in other cases, the purpose of this paper is to present such proof emphasizing the importance of the existence of these rules. By the way, as the ground state of these models can be degenerate as a function of their algebra and, hence, they can support some symmetry-protected topological (SPT) phases, we prove that these non-Abelian fusion rules are always necessary for these SPT phase transitions to occur via a condensation mechanism or/and some global symmetry breaking.
		\end{abstract}
	\end{frontmatter}
	
	\section{A brief overview of the model presented in Ref. \cite{pramod}}
	
		Ref. \cite{pramod} deals with the existence of non-Abelian fusion rules in a class of Abelian models designated by $ H_{N} / \mathds{C} \left( \mathds{Z}_{P} \right) $, whose Hamiltonian operator is
		\begin{equation}
			\mathcal{H} = - \alpha _{A} \sum _{v} A_{v} - \gamma _{C} \sum _{\ell } C_{\ell } \ . \label{h-pramod}
		\end{equation}
		This Hamiltonian, where $ \alpha _{A} $ and $ \gamma _{C} $ are two positive parameters, acts on a Hilbert space
		\begin{equation}
			\underbrace{ \mathfrak{H} _{P} \otimes \ldots \otimes \mathfrak{H} _{P}} _{N_{\ell } \ \textnormal{\tiny{times}}} \ \otimes \ \underbrace{ \mathfrak{H} _{N} \otimes \ldots \otimes \mathfrak{H} _{N}} _{N_{v} \ \textnormal{\tiny{times}}} \ , \label{hilbert-space}
		\end{equation}
		whose states are basically defined once we
		\begin{itemize}
			\item take an \emph{oriented lattice} $ \mathcal{L} _{2} $, with $ N_{v} $ vertices and $ N_{\ell } $ links (which can be occasionally interpreted as one that discretizes some two-dimensional compact orientable manifold $ \mathcal{M} _{2} $ so that there is no need to worry about assigning lattice boundary conditions), and
			\item associate ($ P $- and $ N $-dimensional) Hilbert spaces $ \mathfrak{H} _{P} $ and $ \mathfrak{H} _{N} $ to the $ \ell $-th link and $ v $-th vertex of $ \mathcal{L} _{2} $ respectively, whose vectors can be interpreted as \emph{quantum nits} (qunits), as they are natural generalizations of a quantum bit.
		\end{itemize}
		\begin{figure}[!t]
			\begin{center}
				\begin{adjustbox}{max size={.7\textwidth}{.7\textheight}}
					\begin{tikzpicture}
						% vertices and plaquettes colors
						\draw[color=red!20,fill=red!20] (1,2) rectangle (3,4);
						\draw[color=green!20,fill=green!20] (5.5,4.5) rectangle (8.5,5.5);
						% vertical lines
						\draw[->, color=gray, ultra thick, >=stealth] (0,0) -- (0,2.2);
						\draw[->, color=gray, ultra thick, >=stealth] (0,2) -- (0,4.2);
						\draw[->, color=gray, ultra thick, >=stealth] (0,4) -- (0,6.2);
						\draw[-, color=gray, ultra thick] (0,6) -- (0,8);
						\draw[->, color=gray, ultra thick, >=stealth] (2,0) -- (2,2.2);
						\draw[->, color=gray, ultra thick, >=stealth] (2,2) -- (2,4.2);
						\draw[->, color=gray, ultra thick, >=stealth] (2,4) -- (2,6.2);
						\draw[-, color=gray, ultra thick] (2,6) -- (2,8);
						\draw[->, color=gray, ultra thick, >=stealth] (4,0) -- (4,2.2);
						\draw[->, color=gray, ultra thick, >=stealth] (4,2) -- (4,4.2);
						\draw[->, color=gray, ultra thick, >=stealth] (4,4) -- (4,6.2);
						\draw[-, color=gray, ultra thick] (4,6) -- (4,8);
						\draw[->, color=gray, ultra thick, >=stealth] (6,0) -- (6,2.2);
						\draw[->, color=gray, ultra thick, >=stealth] (6,2) -- (6,4.2);
						\draw[->, color=gray, ultra thick, >=stealth] (6,4) -- (6,6.2);
						\draw[-, color=gray, ultra thick] (6,6) -- (6,8);
						\draw[->, color=gray, ultra thick, >=stealth] (8,0) -- (8,2.2);
						\draw[->, color=gray, ultra thick, >=stealth] (8,2) -- (8,4.2);
						\draw[->, color=gray, ultra thick, >=stealth] (8,4) -- (8,6.2);
						\draw[-, color=gray, ultra thick] (8,6) -- (8,8);
						\draw[->, color=gray, ultra thick, >=stealth] (10,0) -- (10,2.2);
						\draw[->, color=gray, ultra thick, >=stealth] (10,2) -- (10,4.2);
						\draw[->, color=gray, ultra thick, >=stealth] (10,4) -- (10,6.2);
						\draw[-, color=gray, ultra thick] (10,6) -- (10,8);
						% horizontal lines
						\draw[->, color=gray, ultra thick, >=stealth] (-1,1) -- (1.2,1);
						\draw[->, color=gray, ultra thick, >=stealth] (1,1) -- (3.2,1);
						\draw[->, color=gray, ultra thick, >=stealth] (3,1) -- (5.2,1);
						\draw[->, color=gray, ultra thick, >=stealth] (5,1) -- (7.2,1);
						\draw[->, color=gray, ultra thick, >=stealth] (7,1) -- (9.2,1);
						\draw[-, color=gray, ultra thick] (9,1) -- (11,1);
						\draw[->, color=gray, ultra thick, >=stealth] (-1,3) -- (1.2,3);
						\draw[->, color=gray, ultra thick, >=stealth] (1,3) -- (3.2,3);
						\draw[->, color=gray, ultra thick, >=stealth] (3,3) -- (5.2,3);
						\draw[->, color=gray, ultra thick, >=stealth] (5,3) -- (7.2,3);
						\draw[->, color=gray, ultra thick, >=stealth] (7,3) -- (9.2,3);
						\draw[-, color=gray, ultra thick] (9,3) -- (11,3);
						\draw[->, color=gray, ultra thick, >=stealth] (-1,5) -- (1.2,5);
						\draw[->, color=gray, ultra thick, >=stealth] (1,5) -- (3.2,5);
						\draw[->, color=gray, ultra thick, >=stealth] (3,5) -- (5.2,5);
						\draw[->, color=gray, ultra thick, >=stealth] (5,5) -- (7.2,5);
						\draw[->, color=gray, ultra thick, >=stealth] (7,5) -- (9.2,5);
						\draw[-, color=gray, ultra thick] (9,5) -- (11,5);
						\draw[->, color=gray, ultra thick, >=stealth] (-1,7) -- (1.2,7);
						\draw[->, color=gray, ultra thick, >=stealth] (1,7) -- (3.2,7);
						\draw[->, color=gray, ultra thick, >=stealth] (3,7) -- (5.2,7);
						\draw[->, color=gray, ultra thick, >=stealth] (5,7) -- (7.2,7);
						\draw[->, color=gray, ultra thick, >=stealth] (7,7) -- (9.2,7);
						\draw[-, color=gray, ultra thick] (9,7) -- (11,7);
						% vertex vertical line
						\draw[->, ultra thick, >=stealth] (2,1) -- (2,2.2);
						\draw[->, ultra thick, >=stealth] (2,2.0) -- (2,4.2);
						\draw[-, ultra thick] (2,4.0) -- (2,5.0);
						% vertex horizontal line
						\draw[->, ultra thick, >=stealth] (0.0,3) -- (1.2,3);
						\draw[->, ultra thick, >=stealth] (1.0,3) -- (3.2,3);
						\draw[-, ultra thick] (3.0,3) -- (4.0,3);
						% link horizontal line
						\draw[->, ultra thick, >=stealth] (6.0,5) -- (7.2,5);
						\draw[-, ultra thick] (7.0,5) -- (8.0,5);
						% vertices labels
						\draw[color=black,fill=white] (2,3) circle (0.3);
						\node [] (2,3) at (2,3) {$ \gamma $};
						\node [right] (2.1,4) at (2.1,4) {$ a $};
						\node [below] (3,2.9) at (3,2.9) {$ b $};
						\node [left] (1.9,2) at (1.9,2) {$ c $};
						\node [above] (1,3.1) at (1,3.1) {$ d $};
						% link labels
						\draw[color=black,fill=white] (6,5) circle (0.3);
						\node [] (6,5) at (6,5) {$ \alpha $};
						\draw[color=black,fill=white] (8,5) circle (0.3);
						\node [] (8,5) at (8,5) {$ \beta $};
						\node [below] (7.0,4.9) at (7.0,4.9) {$ \ell $};
					\end{tikzpicture}
				\end{adjustbox}
			\end{center}
			\caption{\label{pramod-rede} Piece of an oriented square lattice $ \mathcal{L} _{2} $ that supports the $ H_{N} / \mathds{C} \left( \mathds{Z}_{P} \right) $ models, where we see the rose and light green coloured sectors respectively centred by the $ v $-th vertex and $ \ell $-th link of this lattice. Here, the highlighted links (in black) correspond to Hilbert subspaces in which, for instance, the vertex $ A_{v} $ (the rose-coloured sector) and link $ C_{\ell } $ (the light green coloured sector) operators act effectively. In the case of the links that structure the $ v $-th vertex, they define a subset that we denote by $ S_{v} $.}
		\end{figure}
		In the case of the Hilbert space $ \mathfrak{H} _{P} $, its basis is can be represented by $ \mathcal{B} _{P} = \big\{ \left\vert g \right\rangle : g \in \mathds{Z} _{P} \big\} $, thus justifying the \textquotedblleft Abelian\textquotedblright \hspace*{0.01cm} predicate of this class of models, while the Hilbert space $ \mathfrak{H} _{N} $, has a basis representation given by $ \mathcal{B} _{v} = \big\{ \left\vert \alpha \right\rangle : \alpha \in S \big\} $ where $ S = 1 , 2 , \ldots , N-1 $. The latter can be interpreted as a (left) $ \mathds{C} \left( \mathds{Z}_{P} \right) $\emph{-module} \cite{fulton} when the multiplication (\emph{group action}) $ \mu : \mathds{Z}_{P} \times S \rightarrow S $, which defines how the $ \mathds{Z} _{P} $ group acts on the $ \mathcal{B} _{v} $ elements, is taken into account. This is why these Abelian models were denoted as $ H_{N} / \mathds{C} \left( \mathds{Z}_{P} \right) $. %%%%%%%%%%%%%%%% 87B
		
		Broadly speaking, we can say that the $ H_{N} / \mathds{C} \left( \mathds{Z}_{P} \right) $ models were defined so that they could be interpreted as a subclass of a particular generalization of the \emph{Kitaev Quantum Double Models} $ D \left( G \right) $ \cite{kitaev-1,pachos}: concretely, as a particular generalization where the $ D \left( \mathds{Z} _{P} \right) $ models (i.e., the $ D \left( G \right) $ models with a gauge group $ G = \mathds{Z} _{P} $) were coupled to matter fields on the lattice vertices ($ D_{N} \left( \mathds{Z} _{P} \right) $) \cite{miguel}. Note that, as the operators
		\begin{equation}
			A_{v} = \frac{1}{P} \sum _{g \in G} A^{g} _{v} \ \ \textnormal{and} \ \ C_{\ell } = C_{\ell ,0} \label{h-pramod-operators}
		\end{equation}
		that appear in the Hamiltonian (\ref{h-pramod}) can be identified as the same vertex and link operators of the $ D_{N} \left( \mathds{Z} _{P} \right) $ models, this interpretation makes sense since that (\ref{h-pramod}) is clearly a particular case of the $ D_{N} \left( \mathds{Z} _{P} \right) $ Hamiltonian
		\begin{equation}
			\mathcal{H} _{D_{N} \left( \mathds{Z}_{P} \right) } = - \alpha _{A} \sum _{v} A_{v} - \beta _{B} \sum _{f} B_{f} - \gamma _{C} \sum _{l} C_{l} \label{miguel-hamiltonian}
		\end{equation}
		where $ \beta _{B} = 0 $. Moreover, it is also interesting to note that it is precisely this interpretation the one that explains why $ \mathfrak{H} _{P} $ is considered to model gauge fields. After all, even if the Hamiltonian (\ref{h-pramod}) does not have an operator $ B_{f} = B^{0} _{f} $ that measures the holonomies around the faces of $ \mathcal{L} _{2} $, the definition of the components $ A^{g} _{v} $
		\begin{figure}[!t]
			\begin{center}
				\begin{tikzpicture}[
					scale=0.3,
					equation/.style={thin},
					trans/.style={thin,shorten >=0.5pt,shorten <=0.5pt,>=stealth},
					flecha/.style={thin,->,shorten >=0.5pt,shorten <=0.5pt,>=stealth},
					every transition/.style={thick,draw=black!75,fill=black!20}
					]
					\draw[equation] (-7.5,0.0) -- (-7.5,0.0) node[midway,right] {$ A^{g} _{v} $};
					\draw[trans] (-4.8,2.3) -- (-4.8,-2.3) node[above=2pt,right=-1pt] {};
					\draw[trans] (1.9,2.3) -- (3.1,-0.06) node[above=2pt,right=-1pt] {};
					\draw[trans] (1.9,-2.3) -- (3.1,0.06) node[above=2pt,right=-1pt] {};
					\draw[flecha] (-3.1,0.0) -- (-2.2,0.0) node[above=2pt,right=-1pt] {};
					\draw[flecha] (-2.4,0.0) -- (0.7,0.0) node[above=2pt,right=-1pt] {};
					\draw[trans] (0.2,0.0) -- (1.1,0.0) node[above=2pt,right=-1pt] {};
					\draw[flecha] (-1.0,-1.7) -- (-1.0,-1.0) node[above=2pt,right=-1pt] {};
					\draw[flecha] (-1.0,-1.2) -- (-1.0,1.6) node[above=2pt,right=-1pt] {};
					\draw[trans] (-1.0,1.4) -- (-1.0,1.8) node[above=2pt,right=-1pt] {};
					\draw[trans,fill=white] (-1.0,0.0) circle (0.9);
					\draw[equation] (-1.0,0.0) -- (-1.0,0.0) node[midway] {$ \alpha $};
					\draw[equation] (-1.0,2.4) -- (-1.0,2.4) node[midway] {$ a $};
					\draw[equation] (2.6,0.0) -- (2.6,0.0) node[midway,left] {$ b $};
					\draw[equation] (-1.0,-2.4) -- (-1.0,-2.4) node[midway] {$ c $};
					\draw[equation] (-4.7,0.0) -- (-4.7,0.0) node[midway,right] {$ d $};
					\draw[equation] (4.5,-0.07) -- (4.5,-0.07) node[midway] {$ = $};
					\draw[equation] (6.6,0.15) -- (6.6,0.15) node[midway] {$ \sum $};
					\draw[equation] (6.6,-1.1) -- (6.6,-1.1) node[midway] {$ _{\gamma } $};
					\draw[equation] (11.7,0.1) -- (11.7,0.1) node[midway] {$ \delta \left( \mu \left( g , \alpha \right) , \gamma \right) $};
					\draw[trans] (16.3,2.3) -- (16.3,-2.3) node[above=2pt,right=-1pt] {};
					\draw[trans] (25.6,2.3) -- (26.8,-0.06) node[above=2pt,right=-1pt] {};
					\draw[trans] (25.6,-2.3) -- (26.8,0.06) node[above=2pt,right=-1pt] {};
					\draw[flecha] (19.9,0.0) -- (20.8,0.0) node[above=2pt,right=-1pt] {};
					\draw[flecha] (20.6,0.0) -- (23.7,0.0) node[above=2pt,right=-1pt] {};
					\draw[trans] (23.2,0.0) -- (24.1,0.0) node[above=2pt,right=-1pt] {};
					\draw[flecha] (22.0,-1.8) -- (22.0,-1.0) node[above=2pt,right=-1pt] {};
					\draw[flecha] (22.0,-1.2) -- (22.0,1.6) node[above=2pt,right=-1pt] {};
					\draw[trans] (22.0,1.4) -- (22.0,1.8) node[above=2pt,right=-1pt] {};
					\draw[trans,fill=white] (21.95,0.0) circle (0.9);
					\draw[equation] (21.9,0.0) -- (21.9,0.0) node[midway] {$ \gamma $};
					\draw[equation] (22.1,2.3) -- (22.1,2.3) node[midway] {$ ga $};
					\draw[equation] (26.2,-0.2) -- (26.2,-0.2) node[midway,left] {$ gb $};
					\draw[equation] (22.1,-2.3) -- (22.1,-2.3) node[midway] {$ cg^{-1}  $};
					\draw[equation] (16.4,0.0) -- (16.4,0.0) node[midway,right] {$ dg^{-1}  $};		
				\end{tikzpicture} \\
				\begin{tikzpicture}[
					scale=0.3,
					equation/.style={thin},
					trans/.style={thin,shorten >=0.5pt,shorten <=0.5pt,>=stealth},
					flecha/.style={thin,->,shorten >=0.5pt,shorten <=0.5pt,>=stealth},
					every transition/.style={thick,draw=black!75,fill=black!20}
					]
					\draw[equation] (-8.5,0.0) -- (-8.5,0.0) node[midway,right] {$ C_{\ell , \alpha } $};
					\draw[trans] (-5.6,2.3) -- (-5.6,-2.3) node[above=2pt,right=-1pt] {};
					\draw[trans] (0.1,2.3) -- (1.3,-0.06) node[above=2pt,right=-1pt] {};
					\draw[trans] (0.1,-2.3) -- (1.3,0.06) node[above=2pt,right=-1pt] {};
					\draw[flecha] (-4.4,0.0) -- (-2.1,0.0) node[above=2pt,right=-1pt] {};
					\draw[trans] (-2.3,0.0) -- (-0.2,0.0) node[above=2pt,right=-1pt] {};
					\draw[trans,fill=white] (-4.4,0.0) circle (0.9);
					\draw[trans,fill=white] (-0.2,0.0) circle (0.9);
					\draw[equation] (-2.3,-0.8) -- (-2.3,-0.8) node[midway] {$ a $};
					\draw[equation] (-4.4,0.0) -- (-4.4,0.0) node[midway] {$ \alpha $};
					\draw[equation] (-0.2,-0.02) -- (-0.2,-0.02) node[midway] {$ \beta $};
					\draw[equation] (2.6,-0.07) -- (2.6,-0.07) node[midway] {$ = $};
					\draw[equation] (7.5,0.0) -- (7.5,0.0) node[midway] {$ \delta \left( \mu \left( a , \alpha \right) , \beta \right) $};
					\draw[trans] (12.0,2.3) -- (12.0,-2.3) node[above=2pt,right=-1pt] {};
					\draw[trans] (17.7,2.3) -- (18.9,-0.06) node[above=2pt,right=-1pt] {};
					\draw[trans] (17.7,-2.3) -- (18.9,0.06) node[above=2pt,right=-1pt] {};
					\draw[flecha] (13.2,0.0) -- (15.5,0.0) node[above=2pt,right=-1pt] {};
					\draw[trans] (15.2,0.0) -- (17.4,0.0) node[above=2pt,right=-1pt] {};
					\draw[trans,fill=white] (13.2,0.0) circle (0.9);
					\draw[trans,fill=white] (17.4,0.0) circle (0.9);
					\draw[equation] (15.3,-0.8) -- (15.3,-0.8) node[midway] {$ a $};
					\draw[equation] (13.2,0.0) -- (13.2,0.0) node[midway] {$ \alpha $};
					\draw[equation] (17.4,-0.02) -- (17.4,-0.02) node[midway] {$ \beta $};
				\end{tikzpicture}
			\end{center}
			\caption{\label{pramod-operators-components-definition} Definition of the components $ A^{g} _{v} $ and $ C_{\ell , \alpha } $ in terms of their effective action on $ \mathcal{L} _{2} $, where the group element $ a $ is indexing an $ \left\vert a \right\rangle $ basis element of the Hilbert space $ \mathfrak{H} _{P} $ and the symbol $ \alpha $ indexes an $ \left\vert \alpha \right\rangle $ basis element of the Hilbert space $ \mathfrak{H} _{N} $. Here, $ \delta \left( x , y \right) $ should be interpreted as a Kronecker delta that was written differently for the sake of intelligibility (i.e., $ \delta \left( x , y \right) = \delta _{xy} $).}
		\end{figure}
		shows us that $ A_{v} $ performs the same gauge transformations as the $ D_{N} \left( \mathds{Z}_{P} \right) $ models (see Figure \ref{pramod-operators-components-definition}).
		
		For the sake of completeness, it is also worth noting that the justification for why $ \mathfrak{H} _{P} $ models matter fields follows from Ref. \cite{fradkin}, where lattice gauge theories were coupled to fixed-length scalar (\emph{Higgs}) fields allocated on the lattice vertices. However, it is much more important to pay attention to the fact that the $ H_{N} / \mathds{C} \left( \mathds{Z}_{P} \right) $ models were built as the $ D_{N} \left( \mathds{Z}_{P} \right) $ models without face operators because Ref. \cite{miguel} showed that the presence of such operators in (\ref{miguel-hamiltonian}) causes the magnetic quasiparticles to acquire confinement properties. That is, although it is even possible to transport these magnetic quasiparticles through the lattice, this transport increases the energy of the system and this is not welcome in the context of an eventual quantum computation. %%%%%%%%%%%%%%%% 90A
	
	\section{Some considerations on the vertices and link operators}
	
		Also for the sake of completeness, it is important to point out that if you, the reader, want to know which are, for instance, the ground state degeneracy, the entanglement entropy and the excited states of these lattice models that are defined by using (\ref{h-pramod}), you can consult Ref. \cite{pramod-suggestion}. Fortunately for us, it is not necessary to know these properties in order to present our analysis, which is more qualitative in its philosophy. After all, since Ref. \cite{pramod} shows the emergence of non-Abelian fusion rules in three $ H_{N} / \mathds{C} \left( \mathds{Z}_{P} \right) $ models, but it fails to prove if these rules also exist in other cases, the purpose of this note is to only present such proof emphasizing the importance of the existence of these rules.
		
		As a matter of fact, the best way to evaluate for us to assess whether these rules also exist in these other cases is by noting that, in addition to the matrix representation of the vertex operators being given by
		\begin{equation*}
			A_{v} = \frac{1}{P} \sum _{g \in \mathds{Z} _{P}} M _{v} \left( g \right) \left( \prod _{j^{\prime } \in S^{\uparrow } _{v}} X^{g} _{j^{\prime }} \right) \left( \prod _{j^{\prime \prime } \in S^{\downarrow } _{v}} X^{-g} _{j^{\prime \prime }} \right) \ , \label{a-representation}
		\end{equation*}
		where\footnote{Here, we are considering the same single-qunit computational basis states of Ref. \cite{pramod}, where the vector (ket) $ \left\vert n \right\rangle $, with $ n $ being a natural number, can be represented by a column matrix whose $ n $-th row contains the number $ 1 $ while the others are filled with the number $ 0 $.}
		\begin{itemize}
			\item $ X = \sum _{h \in \mathds{Z} _{P}} \left\vert \left( h + 1 \right) \textnormal{mod} \ P \right\rangle \left\langle h \right\vert $,
			\item $ M \left( g \right) $ is the matrix representation of the \textquotedblleft gauge\textquotedblright \hspace*{0.01cm} group action $ \mu $, and
			\item $ S^{\uparrow } _{v} $ and $ S^{\downarrow } _{v} $ are interpreted as disjoint subsets of $ S_{v} $ such that their link orientations are pointing in and out of the $ v $-th vertex respectively,
		\end{itemize}
		the $ H_{N} / \mathds{C} \left( \mathds{Z}_{P} \right) $ models also have other vertex operators whose matrix representations are
		\begin{equation}
			A_{v,r} = \frac{1}{P} \sum _{g \in \mathds{Z} _{P}} \omega ^{-rg} \cdot M _{v} \left( g \right) \left( \prod _{j^{\prime } \in S^{\uparrow } _{v}} X^{g} _{j^{\prime }} \right) \left( \prod _{j^{\prime \prime } \in S^{\downarrow } _{v}} X^{-g} _{j^{\prime \prime }} \right) \ , \label{a-perp-representation}
		\end{equation}
		where $ r = 1 , 2 , \ldots , P-1 $ and $ \omega = e^{i \left( 2 \pi / P \right) } $ is the generator of the gauge group $ \mathds{Z} _{P} $. A similar comment applies in the case of the existence of other link operators $ C_{\ell ,b} $ that do not define the Hamiltonian (\ref{h-pramod}), which are such that
		\begin{equation*}
			C_{\ell ,s} \left\vert \chi _{v^{\prime }} , \varphi _{\ell } , \chi _{v^{\prime \prime }} \right\rangle = \left\langle \chi _{v^{\prime }} \right\vert \left[ M \bigl( \phi _{\ell } \bigr) \right] ^{-1} \cdot \left( X^{\prime } \right) ^{s} \left\vert \chi _{v^{\prime \prime }} \right\rangle \cdot \left\vert \chi _{v^{\prime }} , \varphi _{\ell } , \chi _{v^{\prime \prime }} \right\rangle \ ,
		\end{equation*}
		where $ s = 1 , 2 , \ldots , N-1 $ and
		\begin{equation*}
			X^{\prime } = \sum _{\lambda \in \mathds{Z} _{N}} \left\vert \left( \lambda + 1 \right) \textnormal{mod} \ N \right\rangle \left\langle \lambda \right\vert \ .
		\end{equation*}
		After all, by taking $ A_{v} = A_{v,0} $, all the operators $ W^{\left( J , K \right) } _{v} $, which are capable of creating quasiparticles by acting on the matter fields, need to be such that
		\begin{equation}
			W^{\left( J , K \right) } _{v} \circ A_{v,0} = A_{v, J} \circ W^{\left( J , K \right) } _{v} \ \ \textnormal{and} \ \ W^{\left( J , K \right) } _{v} \circ C_{\ell ,0} = C_{\ell ,K} \circ W^{\left( J , K \right) } _{v} \ , \label{WAC-relation}
		\end{equation}
		since all the operators $ A_{v,J} $ and $ C_{\ell ,K} $, with $ J \left( K \right) = 1 , 2 , \ldots , P-1 \left( N-1 \right) $, define two complete sets
		\begin{equation*}
			\mathfrak{A} _{v} = \left\{ A_{v,0} ; A_{v,1} ; \ldots ; A_{v,P-1} \right\} \quad \textnormal{and} \quad \mathfrak{C} _{\ell } = \left\{ C_{\ell ,0} ; C_{\ell ,1} ; \ldots ; C_{\ell ,N-1} \right\}
		\end{equation*}
		of orthogonal projectors onto $ \mathfrak{H} _{P} $ and $ \mathfrak{H} _{N} $ respectively. That is, all of these operators that are contained in $ \mathfrak{A} _{v} $ and $ \mathfrak{C} _{\ell } $, which have eigenvalues equal to $ 0 $ and $ 1 $, are such that\footnote{Here, we are assuming (by definition) that $ A_{v,J^{\prime }} \circ C_{\ell ,K^{\prime }} = C_{\ell ,K^{\prime }} \circ A_{v,J^{\prime }} = 0 $ is valid for all the possible values of $ J^{\prime } $ and $ K^{\prime } $ because the $ D_{N} \left( \mathds{Z} _{P} \right) $ model is exactly solvable \cite{miguel}.}
		\begin{equation*}
			A_{v,J^{\prime }} \circ A_{v,J^{\prime \prime }} = A_{v,J^{\prime \prime }} \circ A_{v,J^{\prime }} = 0 \quad \textnormal{and} \quad C_{\ell ,K^{\prime }} \circ C_{\ell ,K^{\prime \prime }} = C_{\ell ,K^{\prime \prime }} \circ C_{\ell ,K^{\prime }} = 0
		\end{equation*}
		hold when $ J^{\prime } \neq J^{\prime \prime } $ and $ K^{\prime } \neq K^{\prime \prime } $, and that
		\begin{equation*}
			\sum ^{N-1} _{J=0} A_{v,J} = \mathbb{1} _{v} \quad \textnormal{and} \quad \sum ^{N-1} _{K=0} C_{\ell ,k} = \mathbb{1} _{\ell } \ .
		\end{equation*}				
		Note that this fact is in full agreement with the Quantum Mechanics requirements \cite{gottfried} because, as any $ H_{N} / \mathds{C} \left( \mathds{Z}_{P} \right) $ vacuum state $ \left\vert \xi _{0} \right\rangle $ needs to be such that
		\begin{equation}
			A_{v,0} \left\vert \xi _{0} \right\rangle = \left\vert \xi _{0} \right\rangle \quad \textnormal{and} \quad C_{\ell ,0} \left\vert \xi _{0} \right\rangle = \left\vert \xi _{0} \right\rangle \label{vacuum-condition}
		\end{equation}
		are valid for all the $ N_{v} $ vertices and $ N_{\ell } $ links that structure $ \mathcal{L} _{2} $, this allows us to decompose the Hilbert space (\ref{hilbert-space}) into the direct sum
		\begin{equation}
			\mathfrak{H} = \mathfrak{H} ^{\left( 0 \right) } \oplus \mathfrak{H} ^{\perp } \ , \label{soma-direta}
		\end{equation}
		where $ \mathfrak{H} ^{\left( 0 \right) } $ and $ \mathfrak{H} ^{\perp } $ are the orthogonal subspaces that contains all the $ H_{N} / \mathds{C} \left( \mathds{Z}_{P} \right) $ vacuum and non-vacuum states respectively. %%%%%%%%%%%%%%%% 89B
		
	\section{\label{why}Why do quasiparticles with non-Abelian fusion rules exist in the $ H_{N} / \mathds{C} \left( \mathds{Z}_{P} \right) $ models?}
	
		However, it is interesting to note that, although Ref. \cite{pramod} did not bother to answer the question that gives name to this Section, it is just its first example (the one where $ \left( N , P \right) = \left( 3 , 2 \right) $) that leads us to the proof that these non-Abelian fusion rules exist in the other $ H_{N} / \mathds{C} \left( \mathds{Z}_{P} \right) $ models. And in order to understand this proof, we need to make three important observations about this $ H_{3} / \mathds{C} \left( \mathds{Z}_{2} \right) $ model and the first one is that, as $ \mathfrak{H} _{N} $ carries an $ N $-dimensional representation of the $ \mathds{C} \left( \mathds{Z}_{P} \right) $ \cite{pramod}, its non-trivial \textquotedblleft gauge\textquotedblright \hspace*{0.01cm} group action can be represented by
		\begin{equation*}
			M \left( 0 \right) = \begin{pmatrix}
				1 & 0 & 0 \\
				0 & 1 & 0 \\
				0 & 0 & 1
			\end{pmatrix} \ \ \textnormal{and} \ \
			M \left( 1 \right) = \begin{pmatrix}
				0 & 1 & 0 \\
				1 & 0 & 0 \\
				0 & 0 & 1
			\end{pmatrix} \ . 
		\end{equation*}
		After all, since this non-trivial representation can also be written as
		\begin{equation}
			M \left( 0 \right) = \begin{pmatrix}
				\mathbb{1} & \mathbf{0} \\
				\mathbf{0} ^{\mathrm{T}} & 1
			\end{pmatrix} \ \ \textnormal{and} \ \
			M \left( 1 \right) = \begin{pmatrix}
				\sigma ^{x} & \mathbf{0} \\
				\mathbf{0} ^{\mathrm{T}} & 1
			\end{pmatrix} \label{action-32}
		\end{equation}
		by taking $ \mathbb{1} $ as the identity matrix of order $ 2 $ and $ \mathbf{0} $ as the zero column matrix, this allows us to recognize that the operators in (\ref{h-pramod-operators}) can be represented by
		\begin{eqnarray*}
			\lefteqn{A_{v} = \frac{1}{2} \left[ M_{v} \left( 0 \right) + M_{v} \left( 1 \right) \prod_{j \in S_{v}} \sigma ^{x} _{j} \right] \ \ \textnormal{and}} \hspace*{11.0cm} \\
			\lefteqn{C_{\ell } = \frac{1}{2} \left[ \begin{pmatrix}
				\mathbb{1} & \mathbf{0} \\
				\mathbf{0} ^{\mathrm{T}} & 0
			\end{pmatrix} _{v^{\prime }} \hspace*{-0.3cm} \otimes \mathbb{1} _{\ell } \otimes \begin{pmatrix}
				\mathbb{1} & \mathbf{0} \\
				\mathbf{0} ^{\mathrm{T}} & 0
			\end{pmatrix} _{v^{\prime \prime }} + \begin{pmatrix}
				\sigma ^{z} & \mathbf{0} \\
				\mathbf{0} ^{\mathrm{T}} & 0
			\end{pmatrix} _{v^{\prime}} \hspace*{-0.3cm} \otimes \sigma ^{z} _{\ell } \otimes \begin{pmatrix}
				\sigma ^{z} & \mathbf{0} \\
				\mathbf{0} ^{\mathrm{T}} & 0
			\end{pmatrix} _{v^{\prime \prime }} \right] } \hspace*{11.0cm} \\
			& & \hspace*{-5.0cm} + \begin{pmatrix}
				\mathbb{0} & \mathbf{0} \\
				\mathbf{0} ^{\mathrm{T}} & 1
			\end{pmatrix} _{v^{\prime }} \hspace*{-0.3cm} \otimes \mathbb{1} _{\ell } \otimes \begin{pmatrix}
				\mathbb{0} & \mathbf{0} \\
				\mathbf{0} ^{\mathrm{T}} & 1
			\end{pmatrix} _{v^{\prime \prime }} \ .
		\end{eqnarray*}
		Here, $ \sigma ^{x} $ and $ \sigma ^{z} $ are the Pauli matrices.
		
		The second important observation is that the $ H_{3} / \mathds{C} \left( \mathds{Z}_{2} \right) $ ground state is algebraically degenerate. After all, as (\ref{action-32}) permutes $ \left\vert 0 \right\rangle _{v} \leftrightarrow \left\vert 1 \right\rangle _{v} $ but fixes $ \left\vert 2 \right\rangle _{v} $ \footnote{Here, we are using the index $ v $ only to emphasize that $ \left\vert \alpha \right\rangle $ is an element associated with a vertex.}, it defines two orbits (one $ 2 $-cycle and one $ 1 $-cycle) \cite{james} and, therefore, there is no transformation, which can be expressed as a product of the operators (\ref{h-pramod-operators}) or/and form a group, that can connect the two $ H_{3} / \mathds{C} \left( \mathds{Z}_{2} \right) $ vacuum states
		\begin{eqnarray}
			\bigl\vert \xi ^{\left( 1 \right) } _{0} \bigr\rangle \negthickspace & = & \negthickspace \prod _{v^{\prime }} A_{v^{\prime }} \left( \bigotimes _{\ell \in \mathcal{L} _{2}} \left\vert 0 \right\rangle \right) \otimes \left( \bigotimes _{v \in \mathcal{L} _{2}} \left\vert 0 \right\rangle \right) \ \ \textnormal{and} \label{ground-state-qdmv-z2z2} \\
			\bigl\vert \xi ^{\left( 2 \right) } _{0} \bigr\rangle \negthickspace & = & \negthickspace \prod _{v^{\prime }} A_{v^{\prime }} \left( \bigotimes _{\ell \in \mathcal{L} _{2}} \left\vert 0 \right\rangle \right) \otimes \left( \bigotimes _{v \in \mathcal{L} _{2}} \left\vert 2 \right\rangle \right) \ . \label{ground-state-qdmv-z2z3-second}
		\end{eqnarray}
		That is, in spite of we are not concerned with any topological aspects of the $ H_{N} / \mathds{C} \left( \mathds{Z}_{P} \right) $ models here, this two-fold degeneracy makes it clear that this $ H_{3} / \mathds{C} \left( \mathds{Z}_{2} \right) $ model has two phases (which characterize each one of these two independent vacuum states (\ref{ground-state-qdmv-z2z2}) and (\ref{ground-state-qdmv-z2z3-second})) that have a kind of algebraic order. We will return to this subject a little later. 

		The third important observation we need to make is that, due to (\ref{action-32}), the matrix representations of $ W^{\left( J , K \right) } _{v} $ are such that
		\begin{equation}
			W^{\left( J , 0 \right) } _{v} = \begin{pmatrix}
				a_{J0} & b_{J0} & c_{J0} \\
				b_{J0} & a_{J0} & c_{J0} \\
				d_{J0} & d_{J0} & r_{J0}
			\end{pmatrix} \ \ \textnormal{and} \ \ W^{\left( J , 1 \right) } _{v} = \begin{pmatrix}
				a_{J1} & b_{J1} & c_{J1} \\
				- b_{J1} & - a_{J1} & - c_{J1} \\
				d_{J1} & - d_{J1} & 0
			\end{pmatrix} \ , \label{W-operators}
		\end{equation}
		whose entries must be interpreted as complex numbers. Because of this, we can already recognize at least two specific operators:
		\begin{itemize}
			\item the first one is
			\begin{equation}
				W^{\left( 0 , 0 \right) } _{v} = \begin{pmatrix}
					1 & 0 & 0 \\	
					0 & 1 & 0 \\
					0 & 0 & 1
				\end{pmatrix} \label{identity-h3z2}
			\end{equation}
			that, being represented by the identity matrix, creates a \emph{vacuum quasiparticle} $ Q^{\left( 0 , 0 \right) } $; and
			\item the second one is
			\begin{equation}
				W^{\left( 1 , 0 \right) } _{v} = \begin{pmatrix}
					0 & 1 & 0 \\
					1 & 0 & 0 \\
					0 & 0 & 1
				\end{pmatrix} \ , \label{one-two-h3z2}
			\end{equation}
			which is exactly the same matrix $ M \left( 1 \right) $ and, therefore, creates a matter excitation $ Q^{\left( 1 , 0 \right) } $ throughout a permutation $ \left\vert 0 \right\rangle _{v} \leftrightarrow \left\vert 1 \right\rangle _{v} $. %%%%%%%%%%%%%%%% 82B
		\end{itemize}
				
		\subsection{Why are these three observations important?}
			
			As naive as it sounds, it is significant to point out that the excitations created by $ W^{\left( 0 , 0 \right) } _{v} $ and $ W^{\left( 1 , 0 \right) } _{v} $ satisfy one of the fundamental requirements that should be satisfied by any quasiparticles: they are such that
			\begin{equation*}
				Q^{\left( 0 , 0 \right) } \times Q^{\left( 1, 0 \right) } = Q^{\left( 1 , 0 \right) } \times Q^{\left( 0 , 0 \right) } \ .
			\end{equation*}
			That is, by changing the order of $ Q^{\left( 0 , 0 \right) } $ and $ Q^{\left( 1 , 0 \right) } $ does not change the result of this fusion. 
			
			Another thing we need to point out here (which may seem far more naive than to say that the fusion between any quasiparticle with a vacuum quasiparticle is commutative) is that $ Q^{\left( 0 , 0 \right) } $ and $ Q^{\left( 1 , 0 \right) } $ are created by permutations that do not involve the element $ \left\vert 2 \right\rangle _{v} $. And in accordance with (\ref{W-operators}), the only operator $ W^{\left( J , 0 \right) } _{v} $ that can replace $ \left\vert 2 \right\rangle _{v} $ by another element, and still leads to a matter excitation $ Q^{\left( 2 , 0 \right) } $ such that
			\begin{equation*}
				Q^{\left( 2 , 0 \right) } \times Q^{\left( J^{\prime } , 0 \right) } = Q^{\left( J^{\prime } , 0 \right) } \times Q^{\left( 2 , 0 \right) }
			\end{equation*}
			where $ J^{\prime } = 0 , 1 $, is
			\begin{equation}
				W^{\left( 2 , 0 \right) } _{v} = \begin{pmatrix}
					0 & 0 & 1 \\
					0 & 0 & 1 \\
					1 & 1 & \mathsf{a}
				\end{pmatrix} \ , \label{non-abelian-operator-h3z2}
			\end{equation}
			where $ \mathsf{a} $ is a complex number. Thus, it is due to the fact that
			\begin{equation}
				W^{\left( 2 , 0 \right) } _{v} \bigl\vert 0 \bigr\rangle _{v} = W^{\left( 2 , 0 \right) } _{v} \bigl\vert 1 \bigr\rangle _{v} = \bigl\vert 2 \bigr\rangle _{v} \ \ \textnormal{and} \ \ W^{\left( 2 , 0 \right) } _{v} \bigl\vert 2 \bigr\rangle _{v} = \bigl\vert 0 \bigr\rangle + \bigl\vert 1 \bigr\rangle _{v} + \mathsf{a} \cdot \bigl\vert 2 \bigr\rangle _{v} \label{non-abelian-condition-h3z2}
			\end{equation}
			that one of the most interesting aspects of the $ H_{3} / \mathds{C} \left( \mathds{Z}_{2} \right) $ model becomes evident. After all, as the composition
			\begin{equation*}
				W^{\left( 2 , 0 \right) } _{v} \circ W^{\left( 2 , 0 \right) } _{v} = \begin{pmatrix}
					1 & 1 & \mathsf{a} \\
					1 & 1 & \mathsf{a} \\
					\mathsf{a} & \mathsf{a} & 2 + \mathsf{a} ^{2}
				\end{pmatrix} = \underbrace{\begin{pmatrix}
					1 & 0 & 0 \\
					0 & 1 & 0 \\
					0 & 0 & 1
				\end{pmatrix}}_{W^{\left( 0 , 0 \right) } _{v}} + \underbrace{\begin{pmatrix}
					0 & 1 & 0 \\
					1 & 0 & 0 \\
					0 & 0 & 1
				\end{pmatrix}}_{W^{\left( 0 , 1 \right) } _{v}} + \ \mathsf{a} \underbrace{\begin{pmatrix}
					0 & 0 & 1 \\
					0 & 0 & 1 \\
					1 & 1 & \mathsf{a}
				\end{pmatrix}}_{W^{\left( 0 , 2 \right) } _{v}}
			\end{equation*}
			is associated with the fusion rule between two excitations $ Q^{\left( 2 , 0 \right) } $, it is clear that this model can support \emph{non-Abelian} fusion rules \cite{pachos}.
			
		\subsection{Can $ Q^{\left( 0 , 0 \right) } $, $ Q^{\left( 1 , 0 \right) } $ and $ Q^{\left( 2 , 0 \right) } $ be interpreted as quasiparticles?}
		
			Although Ref. \cite{pramod} considers the matter excitations (created by manipulating the matter fields) as anyons and put away the face operator from the (\ref{h-pramod}) (in order to try to create favourable conditions for the $ H_{N} / \mathds{C} \left( \mathds{Z}_{P} \right) $ models to be a good candidate to support some kind of quantum computation), the fact is that it is difficult to recognize these matter excitations as quasiparticles. After all, since these matter excitations are not produced in pairs of \textquotedblleft particle\textquotedblright \hspace*{0.01cm} and \textquotedblleft antiparticle\textquotedblright , it is not possible to transport them over $ \mathcal{L} _{2} $ analogous to what happens, for instance, with the $ D \left( G \right) $ quasiparticles (except by a \textquotedblleft teleport\textquotedblright \hspace*{0.01cm} operator
			\begin{equation*}
				W^{\left( J , K \right) } _{v^{\prime \prime }} \circ W^{\left( J , K \right) } _{v^{\prime }}
			\end{equation*}
			that transports it from one vertex $ v^{\prime } $ to another $ v^{\prime \prime } $ completely arbitrary). Yet, once there is nothing that prevents the $ H_{N} / \mathds{C} \left( \mathds{Z}_{P} \right) $ models from serving as a guide for the construction of other lattice models that support these transports, it is essential that
			\begin{equation}
				Q^{\left( J^{\prime } , K^{\prime } \right) } \times Q^{\left( J^{\prime \prime } , K^{\prime \prime } \right) } = Q^{\left( J^{\prime \prime } , K^{\prime \prime } \right) } \times Q^{\left( J^{\prime } , K^{\prime } \right) } \label{basic-rule}
			\end{equation}
			holds for all the possible values of $ \left( J^{\prime } , K^{\prime } \right) $ and $ \left( J^{\prime \prime } , K^{\prime \prime } \right) $ so that the excitations $ Q^{\left( J^{\prime } , K^{\prime } \right) } $ and $ Q^{\left( J^{\prime \prime } , K^{\prime \prime } \right) } $ may be interpreted as quasiparticles. %%%%%%%%%%%%%%%% 82B
					
		\subsection{On the \textquotedblleft absence\textquotedblright \hspace*{0.01cm} of additional quasiparticles}
				
			Since we already know which operators $ W^{\left( J , 0 \right) } _{v} $ are actually capable of creating quasiparticles in the $ H_{3} / \mathds{C} \left( \mathds{Z}_{2} \right) $ model, , it is appropriate to make the same evaluation about $ W^{\left( J , 1 \right) } _{v} $. And one of the first operators we can take in order to make this evaluation is represented by
			\begin{equation}
				W^{\left( 0 , 1 \right) } _{v} = \begin{pmatrix}
					1 & 0 & 0 \\
					0 & -1 & 0 \\
					0 & 0 & 0
				\end{pmatrix} \ . \label{Wz-pauli}
			\end{equation}
			After all, as
			\begin{equation*}
				W^{\left( 0 , 1 \right) } _{v} \bigl\vert 0 \bigr\rangle _{v} = \bigl\vert 0 \bigr\rangle _{v} \ \ \textnormal{and} \ \ W^{\left( 0 , 1 \right) } _{v} \bigl\vert 1 \bigr\rangle _{v} = - \bigl\vert 1 \bigr\rangle _{v} \ ,
			\end{equation*}
			the excitation $ Q^{\left( 0 , 1 \right) } $ that it creates, by acting on the vacuum state (\ref{ground-state-qdmv-z2z2}), behaves effectively as the $ D \left( \mathds{Z} _{2} \right) $ electric quasiparticle\footnote{Note that (\ref{Wz-pauli}) can be rewritten as
			\begin{equation*}
				W^{\left( 0 , 1 \right) } _{v} = \begin{pmatrix}
					\sigma _{z} & \mathbf{0} \\
					\mathbf{0} ^{\mathrm{T}} & 0
				\end{pmatrix}
			\end{equation*}
			where $ \sigma _{z} $ is one of the Pauli matrices \cite{arfken}, what reinforces the behaviour of $ Q^{\left( 0 , 1 \right) } $ as the $ D \left( \mathds{Z} _{2} \right) $ electric quasiparticle. For this reason, it is not wrong to affirm that this is what explains why, for instance, Ref. \cite{pramod} takes the liberty of interpreting these matter excitations as anyons despite the fact that it is not possible to evaluate their statistics. And for this same reason, we will take the same liberty to denote the matter excitations that satisfy the requirement (\ref{basic-rule}) as quasiparticles throughout this Section.}.
				
			Despite what we have just said about $ Q^{\left( 0 , 1 \right) } $ being perfectly correct, there is another result that deserves our attention: it is
			\begin{equation}
				W^{\left( 0 , 1 \right) } _{v} \circ W^{\left( 2 , 0 \right) } _{v} = \begin{pmatrix}
					0 & 0 & 1 \\
					0 & 0 & -1 \\
					0 & 0 & 0
				\end{pmatrix} \neq \begin{pmatrix}
					0 & 0 & 0 \\
					0 & 0 & 0 \\
					1 & -1 & 0
				\end{pmatrix} = W^{\left( 2 , 0 \right) } _{v} \circ W^{\left( 0 , 1 \right) } _{v} \ , \label{option-include}
			\end{equation}
			which shows us that
			\begin{equation*}
				Q^{\left( 0 , 1 \right) } \times Q^{\left( 2 , 0 \right) } \neq Q^{\left( 2 , 0 \right) } \times Q^{\left( 0 , 1 \right) } \ .
			\end{equation*}
			That is, in spite of $ Q^{\left( 0 , 1 \right) } $ behaves effectively as the $ D \left( \mathds{Z} _{2} \right) $ electric quasiparticle, \emph{it cannot be incorporated into a $ H_{3} / \mathds{C} \left( \mathds{Z}_{2} \right) $ model that already admits $ Q^{\left( 2 , 0 \right) } $ as a quasiparticle}. And once this remark extends to other particles $ Q^{\left( J , 1 \right) } $, it is immediate to conclude that the only quasiparticles $ Q^{\left( J , K \right) } $ of this $ H_{3} / \mathds{C} \left( \mathds{Z}_{2} \right) $ model, \emph{which considers $ Q^{\left( 2 , 0 \right) } $ as a quasiparticle}, are those whose fusion rules are given in Table \ref{h-example-32}.
			\begin{table}
				\begin{center}
					\begin{tabular}{|c|ccc|}
						\hline $ \left( J , K \right) $ & $ \left( 0 , 0 \right) $ & $ \left( 1 , 0 \right) $ & $ \left( 2 , 0 \right) $ \\
						\hline $ \left( 0 , 0 \right) $ & $ \left( 0 , 0 \right) $ & $ \left( 1 , 0 \right) $ & $ \left( 2 , 0 \right) $ \\ 
						$ \left( 1 , 0 \right) $ & $ \left( 1 , 0 \right) $ & $ \left( 0 , 0 \right) $ & $ \left( 2 , 0 \right) $ \\ 
						$ \left( 2 , 0 \right) $ & $ \left( 2 , 0 \right) $ & $ \left( 2 , 0 \right) $ & $ \left( 0 , 0 \right) + \left( 1 , 0 \right) + \mathsf{a} \left( 2 , 0 \right) $ \\ 
						\hline 
					\end{tabular}
				\end{center}
					\caption{\label{h-example-32} Fusion rules associated with the quasiparticles $ Q^{\left( J , K \right) } $ of a $ H_{3} / \mathds{C} \left( \mathds{Z}_{2} \right) $ model that considers $ Q^{\left( 2 , 0 \right) } $ as a quasiparticle. Here, each input $ \left( J , K \right) $ corresponds to one of these quasiparticles, which result from a fusion between two quasiparticles that index the rows and columns of this table.}
			\end{table}
			
			Nevertheless, it is worth to note that, although (\ref{option-include}) shows us that the excitation that is created by $ W^{\left( 0 , 1 \right) } _{v} $ does not actually complete a commutative fusion frame with the three quasiparticles in Table \ref{h-example-32}, when we leave $ Q^{\left( 2 , 0 \right) } $ aside another commutative fusion frame can be defined by putting together $ Q^{\left( 0 , 0 \right) } $, $ Q^{\left( 1 , 0 \right) } $ and $ Q^{\left( 0 , 1 \right) } $ with the $ Q^{\left( 1 , 1 \right) } $ that is created by
			\begin{equation*}
				W^{\left( 1 , 1 \right) } _{v} = \begin{pmatrix}
					\sigma _{y} & \mathbf{0} \\
					\mathbf{0} ^{\mathrm{T}} & 0
				\end{pmatrix} \ ,
			\end{equation*}
			where $ \sigma _{y} $ is a Pauli matrix. In other words, this last fusion frame shows that there are two possibilities to define the $ H_{3} / \mathds{C} \left( \mathds{Z}_{2} \right) $ model:
			\begin{itemize}
				\item the first is the one where the two vacuum states (\ref{ground-state-qdmv-z2z2}) and (\ref{ground-state-qdmv-z2z3-second}) can be excited by the operator $ W^{\left( 2 , 0 \right) } _{v} $, which creates a quasiparticle $ Q^{\left( 2 , 0 \right) } $ that exhibits a non-Abelian fusion rule; and
				\item the second is the one where the vacuum state (\ref{ground-state-qdmv-z2z3-second}) can never be excited by the action of some operator $ W^{\left( J , K \right) } _{v} $, whose Abelian fusion rules are shown in Table \ref{h-example-32-alt}.
			\end{itemize}
			\begin{table}%[b]
				\begin{center}
					\begin{tabular}{|c|cccc|}
						\hline $ \left( J , K \right) $ & $ \left( 0 , 0 \right) $ & $ \left( 1 , 0 \right) $ & $ \left( 0 , 1 \right) $ & $ \left( 1 , 1 \right) $ \\
						\hline $ \left( 0 , 0 \right) $ & $ \left( 0 , 0 \right) $ & $ \left( 1 , 0 \right) $ & $ \left( 0 , 1 \right) $ & $ \left( 1 , 1 \right) $ \\ 
						$ \left( 1 , 0 \right) $ & $ \left( 1 , 0 \right) $ & $ \left( 0 , 0 \right) $ & $ \left( 1 , 1 \right) $ & $ \left( 0 , 1 \right) $ \\ 
						$ \left( 0 , 1 \right) $ & $ \left( 0 , 1 \right) $ & $ \left( 1 , 1 \right) $ & $ \left( 0 , 0 \right) $ & $ \left( 1 , 0 \right) $ \\ 
						$ \left( 1 , 1 \right) $ & $ \left( 1 , 1 \right) $ & $ \left( 0 , 1 \right) $ & $ \left( 1 , 0 \right) $ & $ \left( 0 , 0 \right) $ \\ 
						\hline 
					\end{tabular}
				\end{center}
				\caption{\label{h-example-32-alt} Fusion rules associated with the quasiparticles $ Q^{\left( J , K \right) } $ of a $ H_{3} / \mathds{C} \left( \mathds{Z}_{2} \right) $ model that does not consider $ Q^{\left( 2 , 0 \right) } $ as a quasiparticle.}
			\end{table}
			
			However, if we analyse these two possibilities from a physical point of view, we conclude that the second one \emph{may} not make much sense for, at least, two simple reasons. And the first reason is that this capacity, to withdraw the system from its second vacuum state (\ref{ground-state-qdmv-z2z3-second}) by the action of some $ W^{\left( J , K \right) } _{v} $, legitimizes the presence of $ W^{\left( 2 , 0 \right) } _{v} $ among the other operators that create quasiparticles, although it does not appear among those that define the $ H_{3} / \mathds{C} \left( \mathds{Z}_{2} \right) $ Hamiltonian\footnote{Note that, as well as in QFT (where Hamiltonians can be expressed in the \emph{Fock representation} by using the creation $ a^{\dagger } $ and annihilation $ a $ operators \cite{itzykson}), the entire $ D \left( \mathds{Z} _{P} \right) $ energy spectrum can also be well understood from \cite{mf-pedagogical}
			\begin{itemize}
				\item the knowledge of the ground state of these models, and
				\item the excitations created by the action of the operators that compose its Hamiltonian on this ground state.
			\end{itemize}
			Because of this\label{W-action-inclusion}, it becomes valid to recognize $ M_{v} \left( g \right) $ among the operators that create quasiparticles in the $ H_{3} / \mathds{C} \left( \mathds{Z}_{2} \right) $ model, thus justifying the results (\ref{identity-h3z2}) and (\ref{one-two-h3z2}).}. After all, if it were not so, we could not even say that the $ H_{3} / \mathds{C} \left( \mathds{Z}_{2} \right) $ ground state is \textquotedblleft matter degenerated\textquotedblright \hspace*{0.01cm} because, from the matter (fields) point of view, the vacuum state (\ref{ground-state-qdmv-z2z3-second}) would be useless. The second reason is that the fusion rules in Table \ref{h-example-32-alt} are exactly the same as those shown in Table \ref{h-example-22}, which shows the fusion rules associated with the quasiparticles $ Q^{\left( J , K \right) } $ of the $ H_{2} / \mathds{C} \left( \mathds{Z}_{2} \right) $ model. That is, if we define this $ H_{3} / \mathds{C} \left( \mathds{Z}_{2} \right) $ model by making the vacuum state (\ref{ground-state-qdmv-z2z3-second}) useless from the matter (field) point of view, we are dealing with the same $ H_{2} / \mathds{C} \left( \mathds{Z}_{2} \right) $ model\footnote{Although Ref. \cite{pramod} pay attention only to the $ H_{N} / \mathds{C} \left( \mathds{Z}_{P} \right) $ models where $ N > P $, the definition given above includes the case where $ N = P $.}.
			\begin{table}
				\begin{center}
					\begin{tabular}{|c|cccc|}
						\hline $ \left( J , K \right) $ & $ \left( 0 , 0 \right) $ & $ \left( 1 , 0 \right) $ & $ \left( 0 , 1 \right) $ & $ \left( 1 , 1 \right) $ \\
						\hline $ \left( 0 , 0 \right) $ & $ \left( 0 , 0 \right) $ & $ \left( 1 , 0 \right) $ & $ \left( 0 , 1 \right) $ & $ \left( 1 , 1 \right) $ \\ 
						$ \left( 1 , 0 \right) $ & $ \left( 1 , 0 \right) $ & $ \left( 0 , 0 \right) $ & $ \left( 1 , 1 \right) $ & $ \left( 0 , 1 \right) $ \\ 
						$ \left( 0 , 1 \right) $ & $ \left( 0 , 1 \right) $ & $ \left( 1 , 1 \right) $ & $ \left( 0 , 0 \right) $ & $ \left( 1 , 0 \right) $ \\ 
						$ \left( 1 , 1 \right) $ & $ \left( 1 , 1 \right) $ & $ \left( 0 , 1 \right) $ & $ \left( 1 , 0 \right) $ & $ \left( 0 , 0 \right) $ \\ 
						\hline 
					\end{tabular}
				\end{center}
				\caption{\label{h-example-22} Fusion rules associated with the quasiparticles $ Q^{\left( J , K \right) } $ of the $ H_{2} / \mathds{C} \left( \mathds{Z}_{2} \right) $ model, whose ground state is not degenerated. These quasiparticles are created by the operators $ W^{\left( 0 , 0 \right) } _{v} = \mathbb{1} _{v} $, $ W^{\left( 1 , 0 \right) } _{v} = \sigma ^{x} _{v} $, $ W^{\left( 0 , 1 \right) } _{v} = \sigma ^{z} _{v} $ and $ W^{\left( 1 , 1 \right) } _{v} = \sigma ^{y} _{v} $ because its non-trivial action is represented by $ M \left( 1 \right) = \sigma ^{x} $.}
			\end{table} %%%%%%%%%%%%%%%% 90A
			
			Another reason that can be taken into consideration here is based on the results, which were obtained in Ref. \cite{miguel}, for the $ D_{3} \left( \mathds{Z} _{2} \right) $ model (i.e., for the $ H_{3} / \mathds{C} \left( \mathds{Z}_{2} \right) $ model where $ \beta _{B} \neq 0 $). After all, although Ref. \cite{miguel} has not discussed the need to make transitions among the five vacuum states that define the $ D_{3} \left( \mathds{Z} _{2} \right) $ ground state, it makes an interesting observation: it observes that these five vacuum states can be rewritten by using another Hilbert basis, which allows us to recognize that all the lattice vertices have the same matter field $ \left\vert 0 \right\rangle + \left\vert 1 \right\rangle + \mathsf{a} \cdot \left\vert 2 \right\rangle $ with $ \mathsf{a} = 1 $. In this fashion, and in view of the correspondence principle that must be identified between the two classes of $ D_{N} \left( \mathds{Z} _{P} \right) $ and $ H_{N} / \mathds{C} \left( \mathds{Z}_{P} \right) $ models, by noting that $ W^{\left( 2 , 0 \right) } _{v} $ is the only operator that can lead directly to these vacuum states (where all the lattice vertices have the same matter field $ \left\vert 0 \right\rangle + \left\vert 1 \right\rangle + \mathsf{a} \cdot \left\vert 2 \right\rangle $ with $ \mathsf{a} = 1 $), this observation only reinforces the need for this operator to be present in the $ D_{3} \left( \mathds{Z} _{2} \right) $ model and, therefore, in the $ H_{3} / \mathds{C} \left( \mathds{Z}_{2} \right) $ model.
		
		\subsection{\label{dirac-comment} An interesting analogy}
								
			However, it is interesting to note that (\ref{non-abelian-condition-h3z2}) does much more than to make clear that the $ H_{3} / \mathds{C} \left( \mathds{Z}_{2} \right) $ model can be withdrawn from its vacuum state (\ref{ground-state-qdmv-z2z3-second}). What (\ref{non-abelian-condition-h3z2}) does is to show that, if we consider that the vacuum states (\ref{ground-state-qdmv-z2z2}) and (\ref{ground-state-qdmv-z2z3-second}) correspond to two phases that can \emph{coexist} in the same energy regime, it is possible to go from one phase to another, and vice versa, via a \emph{condensation mechanism}. That is, through \label{page-condensation}
			\begin{itemize}
				\item an exchange $ W^{\left( 2 , 0 \right) } _{v} \bigl\vert 0 \bigr\rangle _{v} = \bigl\vert 2 \bigr\rangle _{v} $ on all the lattice vertices for a transition from (\ref{ground-state-qdmv-z2z2}) to (\ref{ground-state-qdmv-z2z3-second}), or
				\item exchanges, which can be carried out by using (several) combinations of the operators $ W^{\left( 1 , 0 \right) } _{v} $ and $ W^{\left( 2 , 0 \right) } _{v} $ that act on all the vertices of $ \mathcal{L} _{2} $, for a transition from (\ref{ground-state-qdmv-z2z3-second}) to (\ref{ground-state-qdmv-z2z2}).
			\end{itemize}
							
			As a matter of fact, this condensation, which can be done by filling all the lattice vertices with quasiparticles, shows us that the vacuum states thus obtained are quite similar, for instance, to the one that was proposed by P. A. M. Dirac in 1929 \cite{dirac-sea}, who claimed that the vacuum could be interpreted as an infinite \textquotedblleft sea\textquotedblright \hspace*{0.01cm} of particles. This Dirac proposal was a rather rudimentary attempt to solve the problem of states with negative energies before the birth of Quantum Electrodynamics \cite{bj-dr,greiner}. Yet, despite this proposal being quite extravagant (by imagining an infinite amount of charges filling the entire space) and does not allow us to calculate anything, it survives today. After all, it intuitively illustrates how to create pairs of particles and antiparticles in the vacuum, although this brings some \textquotedblleft misinterpretation\textquotedblright \hspace*{0.01cm} by considering an electron as a real particle whereas a positron is considered as a hole \cite{halzen}.
				
			Indeed, at the moment that Dirac presented this idea of vacuum to the scientific community, we still did not know all the particles we know today. And it was precisely this lack of knowledge that led him to believe, for instance, that a hole in this \textquotedblleft sea\textquotedblright \hspace*{0.01cm} could be a proton and not a positron, since the last one was also unknown and was only officially discovered by C. D. Anderson in 1932 \cite{positron-discovery} \footnote{Although this official discovery was only recognized in 1932, positrons were first observed in 1929 by D. Skobeltsyn \cite{positron-observation}.}. However, if we analyse this $ H_{3} / \mathds{C} \left( \mathds{Z}_{2} \right) $ model by taking its first vacuum state (\ref{ground-state-qdmv-z2z2}), we see that there is no difference between thinking this quasiparticle $ Q^{\left( 1 , 0 \right) } $ (which has a fusion rule that identifies it as its own anti-quasiparticle) as
			\begin{itemize}
				\item something real, in a situation where $ W^{\left( 1 , 0 \right) } _{v} $ acts on the $ v $-th vertex of a lattice that has all its vertices previously coated by quasiparticles $ Q^{\left( 0 , 0 \right) } $, or
				\item a hole, in a situation where this same $ W^{\left( 1 , 0 \right) } _{v} $ acts on the $ v $-th vertex of a lattice previously filled by quasiparticles $ Q^{\left( 1 , 0 \right) } $.
			\end{itemize}
			In plain English, this $ H_{3} / \mathds{C} \left( \mathds{Z}_{2} \right) $ model \emph{seems} to describe a physical reality as rudimentary as the one that was conjectured by Dirac and others in the early twentieth century. Nevertheless, as this $ H_{3} / \mathds{C} \left( \mathds{Z}_{2} \right) $ model, alone, cannot dictate the general properties of the others, we will return to this issue in the next Section. %%%%%%%%%%%%%%%% 90A
		
		\subsection{\label{general-non-abelian} On the presence of non-Abelian fusion rules in the other cases}
		
			Although we have only proved the existence of the non-Abelian fusion rules in the $ H_{3} / \mathds{C} \left( \mathds{Z}_{2} \right) $ model, it is not hard to generalize this proof to more general $ H_{N} / \mathds{C} \left( \mathds{Z}_{P} \right) $ models where $ N > P \geqslant 2 $. And for this to be done successfully, we need to pay attention to the fact that it is always possible to define these models by representing their \textquotedblleft gauge\textquotedblright \hspace*{0.01cm} group action as
			\begin{equation}
				M \left( g \right) = \begin{pmatrix}
					\mathcal{A} _{1} \left( g \right) & \mathbf{0} \\
					\mathbf{0} ^{\mathrm{T}} & \mathcal{A} _{2} \left( g \right)
				\end{pmatrix} \label{special-action}
			\end{equation}
			where $ \mathcal{A} _{1} $ and $ \mathcal{A} _{2} $ are block diagonal representations of the $ \mathds{Z} _{P} $. Note that this is exactly the situation of the $ H_{3} / \mathds{C} \left( \mathds{Z}_{2} \right) $ model, since (\ref{action-32}) can be recognized as (\ref{special-action}) given that $ \sigma ^{x} $ (i.e., $ X = \sum _{h \in \mathds{Z} _{2}} \left\vert \left( h + 1 \right) \textnormal{mod} \ 2 \right\rangle \left\langle h \right\vert $) generates a faithful representation of the gauge group $ \mathds{Z} _{2} $ and $ 1 $ generates the trivial representation of this same group \cite{james}. Hence, it is clear that a possible representation for the $ H_{N} / \mathds{C} \left( \mathds{Z}_{P} \right) $ \textquotedblleft gauge\textquotedblright \hspace*{0.01cm} group action is
			\begin{equation}
				M \left( g \right) = \begin{pmatrix}
					X^{g} & \mathbf{0} \\
					\mathbf{0} ^{\mathrm{T}} & \mathcal{A} _{2} \left( g \right)
				\end{pmatrix} \ . \label{X-special-action}
			\end{equation}
			
			Anyway, regardless of whether $ \mathcal{A} _{2} \left( g \right) $ is a non-trivial representation of the $ \mathds{Z} _{P} $ or not, the fact is that (\ref{X-special-action}) is defining $ H_{N} / \mathds{C} \left( \mathds{Z}_{P} \right) $ models whose ground states are, at least, two-fold degenerate: after all, while $ X^{g} $ defines one orbit (one $ P $-cycle), the matrix $ \mathcal{A} _{2} \left( g \right) $ is defining, at least, one more. In this way, by recognizing that the only possibility to perform transitions between/among the $ H_{N} / \mathds{C} \left( \mathds{Z}_{P} \right) $ vacuum states is through an operator
			\begin{equation}
				W^{\left( J , 0 \right) } _{v} = \begin{pmatrix}
					B_{J} & C_{J} \\
					C^{\prime } _{J} & D_{J}
				\end{pmatrix} \label{W-general}
			\end{equation}
			where $ C_{J} $ and $ C^{\prime } _{J} $ cannot be zero matrices, it is immediate to conclude that all the matrices that define $ W^{\left( J , 0 \right) } _{v} $ should be such that all commutators
			\begin{equation}
				\left[ \ X^{g} , B_{J} \ \right] \ , \ \ \left[ \ \mathcal{A} _{2} \left( g \right) , D_{J} \ \right] \ , \ \ \left[ \ X^{g} , C_{J} C^{\prime } _{J} \ \right] \ \ \textnormal{and} \ \ \left[ \ \mathcal{A} _{2} \left( g \right) , C^{\prime } _{J} C_{J} \ \right] \label{matrices-commutation}
			\end{equation}
			are equal to zero.
			
			\subsubsection{Non-Abelian fusion rules in the $ H_{N} / \mathds{C} \left( \mathds{Z}_{N-1} \right) $ models}
			
				Note that (\ref{matrices-commutation}) is satisfied by the matrices
				\begin{equation*}
					B_{2} = \begin{pmatrix}
						0 & 0 \\
						0 & 0
					\end{pmatrix} \ , \ \ C_{2} = \begin{pmatrix}
						1 \\
						1 
					\end{pmatrix} \ , \ \ C^{\prime } _{2} = \begin{pmatrix}
						1 & 1 
					\end{pmatrix} \ \ \textnormal{and} \ \ D_{2} = \begin{pmatrix}
						1
					\end{pmatrix}
				\end{equation*}
				that define the operator (\ref{non-abelian-operator-h3z2}), which creates a quasiparticle that exhibits a non-Abelian fusion rule in the $ H_{3} / \mathds{C} \left( \mathds{Z}_{2} \right) $ model. And based on this finding, if we consider a $ H_{N} / \mathds{C} \left( \mathds{Z}_{N-1} \right) $ model whose \textquotedblleft gauge\textquotedblright \hspace*{0.01cm} group action is represented by
				\begin{equation}
					M \left( g \right) = \begin{pmatrix}
						X^{g} & \mathbf{0} \\
						\mathbf{0} ^{\mathrm{T}} & \mathbb{1}
					\end{pmatrix} \ , \label{subspecial-action}
				\end{equation}
				where $ \mathbb{1} $ is the identity matrix of order $ 1 $, it is not hard to see that, by taking
				\begin{itemize}
					\item $ B_{N-1} $ as a zero (square) matrix of order $ N-1 $,
					\item $ C_{N-1} $ as a column matrix that has all its $ N-1 $ rows filled with the number $ 1 $,
					\item $ C^{\prime } _{N-1} $ as the transpose of this matrix $ C_{P} $, and
					\item $ D_{N-1} $ as a (square) matrix of order $ 1 $ whose entry is a complex number $ \mathsf{a} $,
				\end{itemize}
				the operator
				\begin{equation*}
					W^{\left( N-1 , 0 \right) } _{v} = \begin{pmatrix}
						B_{N-1} & C_{N-1} \\
						C^{\prime } _{N-1} & D_{N-1}
					\end{pmatrix}
				\end{equation*}
				creates a quasiparticle that exhibits a non-Abelian fusion because
				\begin{equation}
					W^{\left( N-1 , 0 \right) } _{v} \circ W^{\left( N-1 , 0 \right) } _{v} = \sum _{g \in \mathds{Z} _{N-1}} M_{v} \left( g \right) \ + \ \mathsf{a} \cdot W^{\left( N-1 , 0 \right) } _{v} \ . \label{NN1-fusion-rules}
				\end{equation}
			
			\subsubsection{Non-Abelian fusion rules in the $ H_{N} / \mathds{C} \left( \mathds{Z}_{P} \right) $ models where $ \mathcal{A} _{2} $ is a trivial representation of the $ \mathds{Z}_{P} $}
			
				Similarly, if we consider a $ H_{N} / \mathds{C} \left( \mathds{Z}_{P} \right) $ model whose \textquotedblleft gauge\textquotedblright \hspace*{0.01cm} group action is represented by (\ref{subspecial-action}), but with $ \mathbb{1} $ being an identity matrix of order $ N-P $, it is also not difficult to see that, by considering that
				\begin{itemize}
					\item $ B_{P} $ is a zero (square) matrix of order $ P $,
					\item $ C_{P} $ is a matrix that has all its $ P $ rows and $ N-P$ columns filled with the number $ 1 $,
					\item $ C^{\prime } _{P} $ is the transpose of this matrix $ C_{P} $, and
					\item $ D_{P} $ is a zero (square) matrix of order $ N-P $,
				\end{itemize}
				the operator
				\begin{equation*}
					W^{\left( P , 0 \right) } _{v} = \begin{pmatrix}
						B_{P} & C_{P} \\
						C^{\prime } _{P} & D_{P}
					\end{pmatrix}
				\end{equation*}
				creates a quasiparticle that also exhibits a non-Abelian fusion because
				\begin{eqnarray}
					\lefteqn{W^{\left( P , 0 \right) } _{v} \circ W^{\left( P , 0 \right) } _{v}} \label{NP-fusion-rules} \\
					& = & \negthickspace \left[ 3 - 2 \left( N - P \right) \right] \cdot W^{\left( 0 , 0 \right) } _{v} \ + \ \left( N-P \right) \sum _{g \in \mathds{Z} ^{\ast } _{P}} M_{v} \left( g \right) \ + \ N \hspace*{-0.2cm} \sum _{h \in \mathds{Z} ^{\ast } _{N-P}} \hspace*{-0.2cm} T_{v} \left( h \right) \ . \notag
				\end{eqnarray}
				Here, $ \mathds{Z} ^{\ast } _{R} = \mathds{Z} _{R} \setminus \left\{ 0 \right\} $ and the matrices
				\begin{equation*}
					T \left( h \right) = \begin{pmatrix}
						\mathbb{1} & \mathbf{0} \\
						\mathbf{0} ^{\mathrm{T}} & \left( X^{\prime } \right) ^{h}
					\end{pmatrix} \ ,
				\end{equation*}
				where $ X^{\prime } = \sum _{c \in \mathds{Z} _{N-P}} \left\vert \left( c + 1 \right) \textnormal{mod} \ \left( N - P \right) \right\rangle \left\langle c \right\vert $, represent the operators that are responsible for exciting the vacuum states
				\begin{equation*}
					\bigl\vert \xi ^{\left( k \right) } _{0} \bigr\rangle = \prod _{v^{\prime }} A_{v^{\prime }} \left( \bigotimes _{\ell \in \mathcal{L} _{2}} \left\vert 0 \right\rangle \right) \otimes \left( \bigotimes _{v \in \mathcal{L} _{2}} \left\vert k \right\rangle \right) \ , 
				\end{equation*}
				where $ k = P , \ldots , N-1 $.			
	
	\section{\label{2P-P} What do the $ H_{2P} / \mathds{C} \left( \mathds{Z}_{P} \right) $ models have to tell us about all this?}
			
		Although these results (\ref{NN1-fusion-rules}) and (\ref{NP-fusion-rules}) have already shown that it is possible to define several $ H_{N} / \mathds{C} \left( \mathds{Z}_{P} \right) $ models that exhibit non-Abelian fusion rules, it is also interesting to glance at those models where
		\begin{equation}
			M \left( g \right) = \begin{pmatrix}
				X^{g} & \mathbf{0} \\
				\mathbf{0} ^{\mathrm{T}} & X^{g}
			\end{pmatrix} \ . \label{different-action}
		\end{equation}
		After all, when we take an operator
		\begin{equation*}
			W^{\left( 2P , 0 \right) } _{v} = \begin{pmatrix}
				\mathbf{0} & \sum _{g \in \mathds{Z} _{P}} X^{g} \\
				\sum _{g \in \mathds{Z} _{P}} X^{g} & \mathbf{0}
			\end{pmatrix} \ ,
		\end{equation*}
		whose representation is composed of matrices
		\begin{equation*}
			B_{P} = D_{P} = \mathbf{0} \quad \textnormal{and} \quad C_{P} = C^{\prime } _{P} = \sum _{g \in \mathds{Z} _{P}} X^{g}
		\end{equation*}
		that clearly satisfy (\ref{matrices-commutation}), it is not difficult to see that
		\begin{equation}
			W^{\left( 2P , 0 \right) } _{v} \circ W^{\left( 2P , 0 \right) } _{v} = P \sum _{g \in \mathds{Z} _{P}} M_{v} \left( g \right) \ . \label{2NN-fusion-rules}
		\end{equation}
		Note that the other two examples discussed in Ref. \cite{pramod} (where $ \left( N , P \right) = \left( 4 , 2 \right) $ and $ \left( N , P \right) = \left( 6 , 3 \right) $) are particular cases of the $ H_{2P} / \mathds{C} \left( \mathds{Z}_{P} \right) $ models. %%%%%%%%%%%%%%%% 81B
		
		Nevertheless, despite (\ref{2NN-fusion-rules}) making it clear that these $ H_{2P} / \mathds{C} \left( \mathds{Z}_{P} \right) $ models also support non-Abelian fusion rules, it is precisely this scenario offered by (\ref{different-action}) that allows us to make an interesting remark. And what is this interesting remark? Unlike $ H_{N} / \mathds{C} \left( \mathds{Z}_{P} \right) $ models that are defined by using a \textquotedblleft gauge\textquotedblright \hspace*{0.01cm} group action (\ref{subspecial-action}), where $ \mathbb{1} $ is an identity matrix of order $ N-P $, non-Abelian fusion rules are not necessary for transitions between/among the two vacuum states
		\begin{eqnarray}
			\bigl\vert \xi ^{\left( 1 \right) } _{0} \bigr\rangle \negthickspace & = & \negthickspace \prod _{v^{\prime }} A_{v^{\prime }} \left( \bigotimes _{\ell \in \mathcal{L} _{2}} \left\vert 0 \right\rangle \right) \otimes \left( \bigotimes _{v \in \mathcal{L} _{2}} \left\vert 0 \right\rangle \right) \ \ \textnormal{and} \label{ground-state-qdmv-zp-1} \\
			\bigl\vert \xi ^{\left( 2 \right) } _{0} \bigr\rangle \negthickspace & = & \negthickspace \prod _{v^{\prime }} A_{v^{\prime }} \left( \bigotimes _{\ell \in \mathcal{L} _{2}} \left\vert 0 \right\rangle \right) \otimes \left( \bigotimes _{v \in \mathcal{L} _{2}} \left\vert P \right\rangle \right) \label{ground-state-qdmv-zp-2}
		\end{eqnarray}
		that are defined by (\ref{different-action}). After all, since
		\begin{equation*}
			W^{\left( P , 0 \right) } _{v} = \begin{pmatrix}
				\mathbb{0} & \mathbb{1} \\
				\mathbb{1} & \mathbb{0}
			\end{pmatrix} 
		\end{equation*}
		can connect the two $ P $-cycles defined by (\ref{different-action}), it is not difficult to see that the $ \mathds{Z} _{P} $ global operator
		\begin{equation}
			F = \prod _{v \in \mathcal{L} _{2}} W^{\left( P , 0 \right) } _{v} \label{connection}
		\end{equation}
		is such that
		\begin{equation}
			\bigl\vert \xi ^{\left( 2 \right) } _{0} \bigr\rangle = F \hspace*{0.04cm} \bigl\vert \xi ^{\left( 1 \right) } _{0} \bigr\rangle \ \Leftrightarrow \ \bigl\vert \xi ^{\left( 1 \right) } _{0} \bigr\rangle = F \hspace*{0.04cm} \bigl\vert \xi ^{\left( 2 \right) } _{0} \bigr\rangle \ . \label{connection-zp}
		\end{equation} %%%%%%%%%%%%%%%% 88B

		\subsection{\label{42-example} The $ H_{4} / \mathbb{C} \left( \mathds{Z} _{2} \right) $ as an example}

			By the way, a good example that reinforces this point, and also helps us understand something else about these models, is the $ H_{4} / \mathbb{C} \left( \mathds{Z} _{2} \right) $ model whose \textquotedblleft gauge\textquotedblright \hspace*{0.01cm} group action is represented by (\ref{different-action}) with $ X = \sigma ^{x} $. After all, as its vertex and link operators (\ref{h-pramod-operators}) are expressed as
			\begin{eqnarray*}
				A_{v} \hspace*{-0.2cm} & = & \hspace*{-0.2cm} \frac{1}{2} \left[ M_{v} \left( 0 \right) + M_{v} \left( 1 \right) \prod_{j \in S_{v}} \sigma ^{x} _{j} \right] \ \ \textnormal{and} \\
				C_{\ell } \hspace*{-0.2cm} & = & \hspace*{-0.2cm} \frac{1}{2} \left[ \begin{pmatrix}
					\mathbb{1} & \mathbb{0} \\
					\mathbb{0} & \mathbb{0}
				\end{pmatrix} _{v^{\prime }} \hspace*{-0.3cm} \otimes \mathbb{1} _{\ell } \otimes \begin{pmatrix}
					\mathbb{1} & \mathbb{0} \\
					\mathbb{0} & \mathbb{0}
				\end{pmatrix} _{v^{\prime \prime }} + \begin{pmatrix}
					\sigma ^{z} & \mathbb{0} \\
					\mathbb{0} & \mathbb{0}
				\end{pmatrix} _{v^{\prime}} \hspace*{-0.3cm} \otimes \sigma ^{z} _{\ell } \otimes \begin{pmatrix}
					\sigma ^{z} & \mathbb{0} \\
					\mathbb{0} & \mathbb{0}
				\end{pmatrix} _{v^{\prime \prime }} \right] \\
				& + & \hspace*{-0.2cm} \frac{1}{2} \left[ \begin{pmatrix}
					\mathbb{0} & \mathbb{0} \\
					\mathbb{0} & \mathbb{1}
				\end{pmatrix} _{v^{\prime }} \hspace*{-0.3cm} \otimes \mathbb{1} _{\ell } \otimes \begin{pmatrix}
					\mathbb{0} & \mathbb{0} \\
					\mathbb{0} & \mathbb{1}
				\end{pmatrix} _{v^{\prime \prime }} + \begin{pmatrix}
					\mathbb{0} & \mathbb{0} \\
					\mathbb{0} & \sigma ^{z}
				\end{pmatrix} _{v^{\prime}} \hspace*{-0.3cm} \otimes \sigma ^{z} _{\ell } \otimes \begin{pmatrix}
					\mathbb{0} & \mathbb{0} \\
					\mathbb{0} & \sigma ^{z}
				\end{pmatrix} _{v^{\prime \prime }} \right]
			\end{eqnarray*}			
			respectively\footnote{Here, we are taking $ \mathbb{0} $ as a zero square matrix of order $ 2 $, which is obviously such that $ \mathbb{0} ^{\mathrm{T}} = \mathbb{0} $.}, and its \textquotedblleft gauge\textquotedblright \hspace*{0.01cm} group action permutes $ \left\vert 0 \right\rangle _{v} \leftrightarrow \left\vert 1 \right\rangle _{v} $ and $ \left\vert 2 \right\rangle _{v} \leftrightarrow \left\vert 3 \right\rangle _{v} $, one thing we can already see is that these two orbits (i.e., these two $ 2 $-cycles) \cite{james} define each of the $ H_{4} / \mathbb{C} \left( \mathds{Z} _{2} \right) $ vacuum states (\ref{ground-state-qdmv-zp-1}) and (\ref{ground-state-qdmv-zp-2}) with $ P = 2 $.
		
			Another thing that we can also see from this example is that, as its \textquotedblleft gauge\textquotedblright \hspace*{0.01cm} group action leads to operators $ W^{\left( J , K \right) } _{v} $ that are such that\footnote{As with the matrices (\ref{W-operators}), the entries of (\ref{general-expressions-h4z2-1}) and (\ref{general-expressions-h4z2-2}) must also be interpreted as complex numbers.}
			\begin{eqnarray}
				W^{\left( J , 0 \right) } _{v} \hspace*{-0.2cm} & = & \hspace*{-0.2cm} \begin{pmatrix}
					a_{J0} & b_{J0} & c_{J0} & d_{J0} \\
					b_{J0} & a_{J0} & d_{J0} & c_{J0} \\
					p_{J0} & q_{J0} & r_{J0} & s_{J0} \\
					q_{J0} & p_{J0} & s_{J0} & r_{J0}
				\end{pmatrix} \hspace*{2.0cm} \textnormal{and} \label{general-expressions-h4z2-1} \\
				W^{\left( J , 1 \right) } _{v} \hspace*{-0.2cm} & = & \hspace*{-0.2cm} \begin{pmatrix}
					a_{J1} & b_{J1} & c_{J1} & d_{J1} \\
					- b_{J1} & - a_{J1} & - d_{J1} & - c_{J1} \\
					p_{J1} & q_{J1} & r_{J1} & s_{J1} \\
					- q_{J1} & - p_{J1} & - s_{J1} & - r_{J1}
				\end{pmatrix} \ , \label{general-expressions-h4z2-2}
			\end{eqnarray}
			the operator $ W^{\left( 2 , 0 \right) } _{v} $, which defines the $ \mathds{Z} _{2} $ global operator that connects the two aforementioned vacuum states, creates a quasiparticle $ Q^{\left( 2 , 0 \right) } $ when acting on a single vertex. And given that this conclusion extends to the operator $ W^{\left( P , 0 \right) } _{v} $ that defines the $ \mathds{Z} _{P} $ global operator whatever the values of $ P $, it is not wrong to say that the comment we made at the end of Subsection \ref{dirac-comment} also extends to other $ H_{N} / \mathds{C} \left( \mathds{Z}_{P} \right) $ models: i.e., it is not wrong to say that all these models also \emph{seem} to describe a physical reality as rudimentary as the one that was conjectured by Dirac and others in the early twentieth century.
		
		\subsection{\label{alternative-pv} An alternative point of view}
			
			Note that, as the operators $ A_{v} $ and $ C_{\ell } $ are projectors (i.e., as they are operators that have eigenvalues equal to $ 0 $ and $ 1 $), another point that \emph{seems} to endorse this \textquotedblleft Dirac analogy\textquotedblright \hspace*{0.01cm} is the fact that, when $ \mathcal{L} _{2} $ is an infinite lattice, the lowest energy
			\begin{equation}
				E_{0} = - \alpha _{A} N_{v} - \gamma _{C} N_{\ell } 
			\end{equation}
			of these $ H_{N} / \mathds{C} \left( \mathds{Z}_{P} \right) $ models tends to $ - \infty $. However, there is an interesting observation that we can make within the context, for instance, of Refs. \cite{pramod-suggestion} and \cite{ricardo}, which were published some years after Ref. \cite{pramod}. And in order to understand this interesting observation, it is instructive to continue considering this same $ H_{4} / \mathds{C} \left( \mathds{Z}_{2} \right) $ model, whose \textquotedblleft gauge\textquotedblright \hspace*{0.01cm} group action is represented by (\ref{different-action}) with $ X = \sigma ^{x} $, as an example. After all, by noting that the ground state degeneracy ($ \mathrm{GSD} $) of all these $ H_{N} / \mathds{C} \left( \mathds{Z}_{P} \right) $ models can be determined by
			\begin{equation*}
				\mathrm{GSD} = \mathrm{Tr} \left( \prod _{v \in \mathcal{L} _{2}} A_{v} \prod _{\ell \in \mathcal{L} _{2}} C_{\ell } \right) \ ,
			\end{equation*}
			it is not difficult to demonstrate that, when we analyse this $ H_{4} / \mathds{C} \left( \mathds{Z}_{2} \right) $ model by considering a connected graph\footnote{That is, a topological space that, for instance, includes discretized manifolds as special cases \cite{pramod-suggestion}.} $ \mathcal{X} $, its ground state degeneracy is
			\begin{equation}
				\mathrm{GSD} = \left\vert H^{0} \left( C , G \right) \right\vert = \left\vert \hspace*{0.04cm} \mathrm{Hom} \left( H_{1} \left( \mathcal{X} \right) , H_{2} \left( \mathds{Z} _{2} \right) \right) \hspace*{0.04cm} \right\vert \ . \label{gsd}
			\end{equation} % = 2^{B_{1}} 
			Here, $ C $ and $ G $ are two chain complexes of Abelian groups, and $ H^{0} \left( C , G \right)$ is the $ 0 $-th cohomology of $ G $ with coefficients in $ C $ \footnote{For the sake of completeness, it is interesting to note that these chains $ G $ and $ C $ are related to the group theoretic information and to the graph information of this $ H_{4} / \mathds{C} \left( \mathds{Z}_{2} \right) $ model respectively.} \cite{pramod-suggestion,ricardo}. And this result is very interesting because, since $ \mathcal{L} _{2} $ can be recognized as a closed graph, it is not difficult to prove that, when $ \mathcal{X} $ is taken as an arbitrary closed graph (i.e., an arbitrary graph without \textquotedblleft loose ends\textquotedblright), this $ H_{4} / \mathds{C} \left( \mathds{Z}_{2} \right) $ model has always $ \mathrm{GSD} = 2 $. That is, this $ \mathrm{GSD} $ \textquotedblleft coincides\textquotedblright \hspace*{0.01cm} with the number of vacuum states (\ref{ground-state-qdmv-zp-1}) and (\ref{ground-state-qdmv-zp-2}), and this \textquotedblleft coincidence\textquotedblright \hspace*{0.01cm} is not the result of chance: it results from the fact that, although the Refs. \cite{pramod}, \cite{pramod-suggestion} and \cite{ricardo} explore different contexts, all they need to converge in several points due to the correspondence principle that needs to be identified between/among their models.
			
			By the way, given this correspondence principle, it is also important to observe that, when $ \mathcal{X} $ is open with edges or has \textquotedblleft loose ends\textquotedblright , this same result (\ref{gsd}) indicates that the $ \mathrm{GSD} $ increases due to the existence of the gapless edge states. And why is this important to observe? Because, in the case of these gapless edge states, they correspond precisely to the two $ \mathds{Z} _{2} \times \mathds{Z} _{2} $ symmetry-protected topological (SPT) phases, which originate in the same two $ 2 $-cycles that are defined by the \textquotedblleft gauge\textquotedblright \hspace*{0.01cm} group action of this $ H_{4} / \mathds{C} \left( \mathds{Z}_{2} \right) $ model. Thus, as all these aforementioned results must be valid with respect to this correspondence principle (which needs to be recognized between this graph analysis and the one we present here), it is not wrong to conclude that this model support two $ \mathds{Z} _{2} \times \mathds{Z} _{2} $ SPT phases (\ref{ground-state-qdmv-zp-1}) and (\ref{ground-state-qdmv-zp-2}), and that the transitions between them occur due to a $ \mathds{Z}_{2} $ global symmetry breaking. %%%%%%%%%%%%%%%% 92A
		
			Note that a similar thing happens in the $ H_{3} / \mathds{C} \left( \mathds{Z}_{2} \right) $ model which, when analysed from the point of view of Refs. \cite{pramod-suggestion} and \cite{ricardo}, also has $ \mathrm{GSD} = 1 + 2^{\mathrm{B} _{1}} $ when $ \mathcal{X} $ is an arbitrary closed graph. Here, $ \mathrm{B} _{1} $ is the first Betti number \cite{hatcher}. Nevertheless, although it is also possible to assert, for instance, that the $ 2 $-cycle defined by the \textquotedblleft gauge\textquotedblright \hspace*{0.01cm} group action (\ref{action-32}) allows us to interpret (\ref{ground-state-qdmv-z2z2}) as a $ \mathds{Z} _{2} \times \mathds{Z} _{2} $ SPT phase, no global operator, which connects (\ref{ground-state-qdmv-z2z2}) and (\ref{ground-state-qdmv-z2z3-second}) in the same sense of (\ref{connection}), can generate a group. In this fashion, by noting that this impossibility is directly related to the fact that this global operator needs to connect two $ k^{\prime } $- and $ k^{\prime \prime } $-cycles of different degrees (i.e, where $ k^{\prime } \neq k^{\prime \prime } $), this suggests that the transitions between this $ \mathds{Z} _{2} \times \mathds{Z} _{2} $ SPT phase and the one that follows from the $ 1 $-cycle (which seems to have the graph topological order discussed in Ref. \cite{pramod-suggestion}) occur due to another kind of symmetry breaking.
		
		\subsection{\label{conflicts} Does this symmetry breaking interpretation conflict with the possible interpretation of the $ H_{3} / \mathds{C} \left( \mathds{Z}_{2} \right) $ vacuum states as Dirac \textquotedblleft seas\textquotedblright ?}
		
			Since we have just said that these phase transitions occur due to some symmetry breaking, one thing that you, the reader, might be wondering is: how does this symmetry breaking interpretation reconcile with the one, which we presented in Section \ref{dirac-comment}, where each of the $ H_{3} / \mathds{C} \left( \mathds{Z}_{2} \right) $ vacuum states was interpreted as a kind of analogue of the Dirac \textquotedblleft sea\textquotedblright ? And if you are wondering about this, the only thing we ask you to observe is that, whenever we saw this analogy, we used the term \emph{\textquotedblleft seem(s)\textquotedblright }. And why did we do it? We did this because the truth is that, if these $ H_{N} / \mathds{C} \left( \mathds{Z}_{P} \right) $ models describe some physical reality, it is still not very clear what physical reality is.
			
			Of course, as these $ H_{N} / \mathds{C} \left( \mathds{Z}_{P} \right) $ models define a subclass of the $ D_{N} \left( \mathds{Z}_{P} \right) $ models, it allows us to infer, for instance, that the quasiparticles $ Q^{\left( J , K \right) } $ have electrical properties because they fuse with the electric quasiparticles inherited from the $ D_{N} \left( \mathds{Z}_{P} \right) $ models. And another fact that reinforces this inference is that, when the \textquotedblleft gauge\textquotedblright \hspace*{0.01cm} group action is not trivial, it allows us to recognize an electrostatic interaction between/among, at least, quasiparticles $ Q^{\left( J , K \right) } $ that have the same flavour (i.e., that have the same $ \left( J , K \right) $ index): after all, when we have only two quasiparticles $ Q^{\left( J , K \right) } $, with the same flavour $ \left( J , K \right) $, on two vertices $ v^{\prime } $ and $ v^{\prime \prime } $ of $ \mathcal{L} _{2} $, it is not difficult to see that the energy of this system is equal to
			\begin{itemize}
				\item $ E_{0} + \gamma _{C} \left( n_{v^{\prime }} + n_{v^{\prime \prime }} \right) -1 $, when $ v^{\prime } $ and $ v^{\prime \prime } $ are neighbours, and
				\item $ E_{0} + \gamma _{C} \left( n_{v^{\prime }} + n_{v^{\prime \prime }} \right) $, otherwise.
			\end{itemize}
			Here, $ n_{v^{\prime }} $ and $ n_{v^{\prime \prime }} $ are positive real numbers that denote the number of links that share $ v^{\prime } $ and $ v^{\prime \prime } $ respectively. But the fact is that, although we are capable of inferring some things about $ Q^{\left( J , K \right) } $, these $ H_{N} / \mathds{C} \left( \mathds{Z}_{P} \right) $ models describe a physical reality that is quite rudimentary, which does not even allow us to know, for instance, what the spin of these quasiparticles is. In this way, as these $ H_{N} / \mathds{C} \left( \mathds{Z}_{P} \right) $ models do not allow us to make a deeper comparison between their vacuum states and the Dirac \textquotedblleft sea\textquotedblright , we prefer to be cautious and, for now, just point to this playful analogy. After all, all the vacuum states that arise from the \textquotedblleft gauge\textquotedblright \hspace*{0.01cm} group action are, in fact, defined by filling all the lattice vertices with quasiparticles $ Q^{\left( J , K \right) } $ that have the same flavour.
			
			In any case, it is interesting to note that there are several works that, for instance, explore this same analogy more deeply in some contexts that do not seem to be so distant from the $ D_{N} \left( \mathds{Z}_{P} \right) $ models \cite{amb,zohar,shankar,julian,maroncelli}. And such works make it clear that there is nothing wrong with interpreting the vacuum states of some lattice models, especially those that can be interpreted as lattice gauge theories, like Dirac \textquotedblleft seas\textquotedblright . Nevertheless, as Dirac claimed that the vacuum could be interpreted as an infinite \textquotedblleft sea\textquotedblright \hspace*{0.01cm} of particles and, therefore, a deeper analogy requires us to explain, for instance, how is it physically possible to go from one phase to another when, by only using $ F $, we need to cross an infinite energy barrier in the thermodynamic limit, it becomes plausible to say that these phase transitions occur due to a global symmetry breaking because it is not wrong. %%%%%%%%%%%%%%%% 89A
			
	\section{\label{remarks} Final remarks}
	
		In accordance with what we just saw, it is quite clear that Ref. \cite{pramod} is correct in stating that the $ H_{N} / \mathds{C} \left( \mathds{Z}_{P} \right) $ models support the presence of matter excitations $ Q^{\left( J , K \right) } $ that exhibit non-Abelian fusion rules. After all,
		\begin{itemize}
			\item these non-Abelian fusion rules can always be identified when the \textquotedblleft gauge\textquotedblright \linebreak group action is represented by (\ref{special-action}), and
			\item when $ \mathcal{A} _{1} \left( g \right) $ and $ \mathcal{A} _{2} \left( g \right) $ define $ k $-cycles of different degrees, these non-Abelian fusion rules are always necessary for (phase) transitions between/\linebreak among the $ H_{N} / \mathds{C} \left( \mathds{Z}_{P} \right) $ vacuum states.
		\end{itemize}
		Note that, although we have identified an interesting analogy between the $ H_{N} / \mathds{C} \left( \mathds{Z}_{P} \right) $ vacuum states and Dirac \textquotedblleft seas\textquotedblright , this playful analogy should not be taken too seriously a priori. After all, even though it is not impossible to explore/investigate this analogy more deeply (perhaps by trying to bring, for instance, the $ D_{N} \left( \mathds{Z}_{P} \right) $ models closer to those in Refs. \cite{amb}, \cite{zohar}, \cite{shankar}, \cite{julian} and \cite{maroncelli}), this analogy was only explored to reinforce to you, the reader, that all the $ H_{N} / \mathds{C} \left( \mathds{Z}_{P} \right) $ vacuum states, which arise from the \textquotedblleft gauge\textquotedblright \hspace*{0.01cm} group action, can be obtained by filling all the lattice vertices (i.e., via a condensation procedure) with quasiparticles $ Q^{\left( J , K \right) } $ that have the same flavour. And although we have said, for instance, that $ F $ is a $ \mathds{Z} _{P} $ global operator, which connects the two $ H_{2P} / \mathds{C} \left( \mathds{Z}_{P} \right) $ vacuum states (\ref{ground-state-qdmv-zp-1}) and (\ref{ground-state-qdmv-zp-2}) (i.e., the two $ \mathds{Z} _{2P} \times \mathds{Z} _{2P} $ SPT phases) by performing a $ \mathds{Z}_{P} $ global symmetry breaking, it is quite clear that this does not contradict this playful analogy because $ W^{\left( P , 0 \right) } _{v} $ actually creates a quasiparticle on the lattice vertices it acts on.
		
		Anyway, another thing that is important to mention here is that, despite Ref. \cite{pramod} has identified several operators that produce matter excitations, it is necessary to be a little careful before saying that all these excitations can be interpreted as quasiparticles in the $ H_{N} / \mathds{C} \left( \mathds{Z}_{P} \right) $ models. After all, in accordance with the result (\ref{option-include}), it is clear that not all these matter excitations satisfy the requirement (\ref{basic-rule}), which is critical for these excitations to be classified as quasiparticles. Nonetheless, by noting that the $ H_{N} / \mathds{C} \left( \mathds{Z}_{P} \right) $ models can be identified as a subclass of the $ D_{N} \left( \mathds{Z} _{P} \right) $ models, it is immediate to conclude that these non-Abelian fusion rules can also be observed in these $ D_{N} \left( \mathds{Z} _{P} \right) $ models.		
	
		By the way, despite these non-Abelian fusion rules are quite similar to those of the Fibonacci anyons \cite{pachos}, all lead us to believe that they (still) \emph{may} not be used to perform any kind of quantum computation with these $ H_{N} / \mathds{C} \left( \mathds{Z}_{P} \right) $ models. After all, the possibility to perform a quantum computation with the $ D \left( G \right) $ models (which clearly serve as the foundation for the $ H_{N} / \mathds{C} \left( \mathds{Z}_{P} \right) $ models) is linked to the possibility of defining braids \cite{braids,bonesteel}, which are formed in $ \mathcal{M} _{2} \times \left[ 0 , 1 \right] $ due to the transport of quasiparticles in $ \mathcal{M} _{2} $. Nevertheless, it is interesting to note that, although it does not \emph{seem} to be possible to transport matter excitations $ Q^{\left( J , K \right) } $ through the lattice (and, therefore, to evaluate their statistics), there are good indicators that they can be interpreted as quasiparticles: one of them is, for example, the fact of the \textquotedblleft gauge\textquotedblright \hspace*{0.01cm} group action allows us to recognize that these matter excitations behave effectively as the $ D \left( \mathds{Z} _{P} \right) $ electric quasiparticles. Wherefore, as Ref. \cite{mf-pedagogical} seems to point to the possibility of building a correspondence principle between the three-dimensional $ D \left( \mathds{Z}_{P} \right) $ models and the two-dimensional $ D_{N} \left( \mathds{Z} _{P} \right) $ models, it seems important to evaluate whether these matter excitations can be transported in the three-dimensional $ H_{N} / \mathds{C} \left( \mathds{Z}_{P} \right) $ models. This will be evaluated in a future work. %%%%%%%%%%%%%%%% 90A
		
	\section*{Acknowledgments}
	
		This work has been supported by CAPES (ProEx) and CNPq (grant 162117/ \linebreak 2015-9). We thank J. L. M. Assirati, U. A. Maciel Neto and D. V. Tausk for some mathematical discussions, as well as G. T. Landi and P. Teotonio Sobrinho for some physical discussions, on subjects concerning this project. Special thanks are also due to the reviewer of this paper, who, although we do not know his name, made some relevant remarks that, for instance, led us to write Subsections \ref{42-example}, \ref{alternative-pv} and \ref{conflicts}. In particular, M. F. also thanks F. Diacenco Xavier for friendly support during this work. This work is dedicated to A. B. Chapisco. %%%%%%%%%%%%%%%% 89 Legendas

\end{document}